\documentclass[11pt]{article}

\usepackage{enumerate}

\usepackage{a4wide}
\usepackage{footnpag,nicefrac,comment}
\usepackage{graphicx}

\usepackage{amsmath}
\usepackage{amsfonts}
\usepackage{amssymb}
\usepackage{color}

\newcommand{\eg}{\emph{eg.} }
\newcommand{\refe}[1]{(\ref{#1})}
\newcommand{\tr}{\; \text{Tr} \;}

\newcommand{\bnormal}[1]{{}^{{}_\star}_{{}^\star} #1 {}^{{}_\star}_{{}^\star}}

\newcommand{\eqna}[1]{\begin{eqnarray} #1 \end{eqnarray}}
\newcommand{\eqali}[1]{\begin{align} #1 \end{align}}

\newcommand{\comm}[2]{\left[ #1 , #2 \right]}

\newcommand{\im}{\mathfrak{Im}}
\newcommand{\re}{\mathfrak{Re}}
\newcommand{\corr}[1]{\left< #1 \right>}
\newcommand{\parent}[1]{\left(#1\right)}

\newcommand{\module}[1]{\left\vert #1 \right\vert}

\newcommand{\sing}[1]{\left\{ #1 \right\}}
\newcommand{\croch}[1]{\left[ #1 \right]}
\newcommand{\bbar}[1]{\left| #1 \right|}

\newcommand{\zamcorr}[2]{\parent{\left. #1 \right| #2 }}
\newcommand{\Gam}{{\bf \Gamma}}

\newcommand{\symform}[2]{\bbar{\begin{array}{c} #1 \\ #2 \end{array}}}

\newcommand{\asymform}[2]{\parent{\begin{array}{c} #1 \\ #2 \end{array}}}
\newcommand{\Asymform}[2]{\croch{\begin{array}{c} #1 \\ #2 \end{array}}}

\long\def\symbolfootnote[#1]#2{\begingroup%
\def\thefootnote{\fnsymbol{footnote}}\footnote[#1]{#2}\endgroup}
\def\di{\mathrm{d}}

\begin{document}

\begin{titlepage}

\vskip 2.5cm

\centerline{\LARGE Rolling tachyons for separated brane-antibrane systems }

\vskip 1.6cm
\centerline{\bf Dan~Isra\"el$^\clubsuit$ and Flavien Kiefer$^{\clubsuit,\spadesuit}$\symbolfootnote[2]{Email:
israel@iap.fr,kiefer@iap.fr}}
\vskip 0.5cm
\

\vskip 0.3cm

\vskip 0.3cm
\centerline{\sl $^\clubsuit$Institut d'Astrophysique de Paris,
98bis Bd Arago, 75014 Paris, France\footnote{Unit\'e mixte de Recherche
7095, CNRS -- Universit\'e Pierre et Marie Curie}
}

\vskip 0.3cm
\centerline{\sl $^\spadesuit$
LPTENS, \'Ecole Normale Sup\'eieure, 
24 rue Lhomond, 75231 Paris cedex 05, France\footnote{Unit\'e mixte de Recherche 
du CNRS et de l'\'Ecole Normale Sup\'erieure associ\'ee \`a l'Universit\'e Pierre et Marie Curie, UMR 8549.}}

\vskip 1.4cm

\centerline{\bf Abstract} \vskip 0.5cm  
We consider tachyon condensation between a D-brane and an anti-D-brane in superstring theory, when they are separated 
in their common transverse directions. A simple rolling tachyon solution, that describes the time evolution of the process, 
is studied from the point of view of boundary conformal field theory. By computing the boundary beta-functions of the system, one 
finds that this theory is conformal, hence corresponds to an exact solution of the string theory 
equations of motion. By contrast, the time-reversal-symmetric rolling tachyon is not conformal. These results put constraints on the  
space-time effective actions for the system.

\noindent

\vfill

\end{titlepage}

\section{Introduction}
Annihilation of D-branes of opposite Ramond-Ramond charge is one of the fundamental processes of string theory. Tachyon condensation on brane-antibrane systems has 
also important cosmological applications, either as a tractable model of a time-dependent process in string theory, or concretely in D-brane inflation 
models~\cite{Alexander:2001ks}. It also appears in holographic models of QCD, to describe chiral symmetry breaking~\cite{Sakai:2004cn}.

Whenever the distance  between the branes is smaller than the critical value $r_c$, the ground state in the  brane-antibrane open 
string sectors becomes tachyonic. It has been conjectured long ago that the condensation of this complex-valued tachyon leads to the closed string vacuum, 
corresponding to the minimum of the tachyon potential~\cite{Sen:1998sm}, and partially confirmed by string field theory computations~\cite{Berkovits:2000zj}.

In the case where the brane and the anti-brane are coincident in their common transverse directions, this system has been thoroughly studied using 
background-independent string field theory (BSFT)~\cite{Witten:1992qy,Witten:1992cr,Shatashvili:1993kk}. In this approach, one considers the two-dimensional worldsheet 
conformal field theory on the disk with marginal and relevant boundary perturbations.  It allows to compute the exact off-shell 
tree-level tachyon potential~\cite{Kutasov:2000qp,Kutasov:2000aq}.

On-shell configurations corresponding to real-time tachyon condensation on unstable D-branes are also of interest, especially whenever the boundary 
conformal field theory (BCFT) is known. For unstable D-branes, a first type of solution, known as the {\it full S-brane} was found by Sen and represents 
a time-reversal symmetric process~\cite{Sen:2002nu}. The second type of solution, known as the {\it half S-brane}~\cite{Karczmarek:2003xm,Gutperle:2003xf}, 
represents the more realistic case of a tachyon starting, from $t\to -\infty$, at the maximum of its potential. It is straightforward to extend these results 
to coincident brane-antibrane pairs. 

Although the gradient of the tachyon field on the rolling tachyon solutions is very large, it should make sense to consider a spacetime effective action that describes slowly 
varying  perturbations thereof. Remarkably, as was shown by Kutasov and Niarchos~\cite{Kutasov:2003er}, it is possible to find {\it unambiguously} 
the effective action for the tachyon and its first derivative asking only that {\it (i)} the rolling tachyon discussed above is a solution to its equations of motion and that 
{\it (ii)} the on-shell Lagrangian on this solution is equal to the disk partition function with the time-like zero mode unintegrated. 
Upon a simple field redefinition, it coincides also with the "tachyon-DBI" action that was earlier proposed by Garousi~\cite{Garousi:2000tr},\footnote{
RR-couplings were added to this action in~\cite{Garousi:2007fk,Garousi:2008ge}.} and is able to reproduce 
correctly N-point tachyon amplitudes~\cite{Niarchos:2004rw}.\footnote{A different an interesting approach to tachyon effective actions on brane-antibrane pairs 
was given  in~\cite{Erkal:2009xq}.}

Surprisingly, not much of this program has been carried out for the system of a D-brane and an anti-D-brane at finite distance --~letting aside the even more interesting and 
challenging case  of brane-antibrane scattering. The brane separation is a modulus at tree-level, even though a brane-antibrane potential is generated at one 
string loop~\cite{Banks:1995ch}. Hence, we can ask whether tachyon condensation at fixed separation is possible.  One may expect  different spacetime physics compared to the 
coincident case, especially in the limit where the absolute value of the tachyon mass is small in string units.

With cubic string field theory, 
an approximation of the tachyon potential as a function of the fixed brane-antibrane separation $r$ has been computed few years ago using level 
truncation at next-to-leading order in~\cite{Bagchi:2008et}. In BSFT, the framework for studying the T-dual configuration --~a brane-antibrane pair 
compactified along a worldvolume direction, with a relative Wilson line~-- was set in the works~\cite{Kraus:2000nj} and~\cite{Takayanagi:2000rz}. There, the worldsheet action of the 
system, including the background spacetime gauge fields along  with the complex tachyon, was set. 
Unfortunately, the Abelian gauge field T-dual to $r$ was set to zero 
in order to simplify the path integral computation.\footnote{Using these results, the spacetime effective action with non-zero gauge field profiles 
was conjectured in~\cite{Jones:2002sia} as a plausible covariantization, however it was not derived from first principles.}

Finally, the 'half S-brane' rolling tachyon solution describing condensation at fixed, finite distance is not really understood, let alone the effective action of which it 
should be a solution. In~\cite{Bagchi:2008et} this problem was studied using conformal perturbation theory, which is expected to be valid, in spacetime terms, for 
very early times at the onset of tachyon condensation. Surprisingly, it was found that the boundary interaction corresponding to the rolling tachyon ceases to be marginal 
for a countable set of values of $|r|$ larger than $r_c/\sqrt{2}$.

In this note, we show that, taking in particular into account the effect of contact terms that are dictated by worldsheet supersymmetry, the rolling tachyon 
boundary interaction seems to be  exactly marginal for all values of $|r|$ below the critical separation. Study of beta-functions for the system illuminates 
the crucial role of the contact term. The latter is able to cancel the power-like short distance singularity that arises at second order in perturbation 
theory for $|r|>1/2$. At fourth order, it cancels all but one power-like singularity that is present for $|r|>\sqrt{7}/4$, for which an higher-order contact 
term is needed. Nevertheless, the potentially dangerous logarithmic singularities, that could occur for  certain values of $|r|$, vanish by themselves without the help of 
the contact term. 

We  find that the beta-functions of the theory  are zero to all orders for $|r|<\sqrt{17}/6$, while for larger values of $|r|$ they vanish at least up to 
order five in perturbation theory. Thus, we expect that the perturbative expansion  in the boundary tachyon perturbations does not break conformal invariance on 
the boundary, for any sub-critical separation.

Unexpectedly, we find that the 'full S-brane' rolling tachyon is not a boundary conformal field theory, for any non-zero separation between the branes. 
In that case the beta-function for the distance-changing boundary operator does not vanish. It implies that the corresponding space-time tachyon profile is 
not a solution of the equations of motion. It seems nevertheless that a more general solution than the 'half S-brane' exists, for which the tachyon starts from and 
comes back to the tachyon vacuum; its physical meaning is not obvious though, since the phase of the complex tachyon cannot stay constant.

From these results we learn that there should exist a space-time effective action for the system, that is valid for any $0\leqslant |r| <r_c$ (to be more precise, 
the effective action for the tachyon and distance field should admit a solution where the distance is a constant). Effective actions were proposed in  
the past by Sen~\cite{Sen:2003tm} and Garousi~\cite{Garousi:2004rd}. However its domain of validity  is not clear. 
Indeed, it does not allow as a solution a tachyon  condensation at fixed distance, even in the regime of small brane separation in string units.  

Imposing the existence of the 'half S-brane' solution at fixed distance fixes the effective action up to second order 
in the tachyon  field. In order to get the fully explicit effective action around this rolling tachyon at fixed distance without further hypothesis, we can proceed as 
in~\cite{Kutasov:2003er} and try to fix all the coefficients of a generic first-order Lagrangian expressed in power series. It fails to give a single answer for two reasons. 
First, as the 'full S-brane' solution seems not be allowed, the constraints from the tachyon equations of motion are weaker. Second, we would need to compute the disk partition 
function, to all orders in the tachyon coupling; for a generic distance analytical results for the perturbative integrals seem out of reach, from the fourth order.

This work is organized as follows. In section~\ref{sec:braneaction} we give some background on the brane-antibrane worldsheet action on the disk, emphasizing the role of the Fermi 
multiplets that realize the Chan-Patton degrees of freedom. In section~\ref{sec:contactterms} we discuss the role of contact terms in canceling the divergences that arises when 
tachyon perturbations collide. In section~\ref{sec:secbeta} we examine the system from the point of view of boundary renormalization group flow, and obtain our main results 
about the marginality of the rolling tachyon profile. Finally in the discussion we give the implications of our results for space-time effective actions. Some 
lengthy computations are given in the appendices. 

\section{Brane-antibrane worldsheet action}
\label{sec:braneaction}
In this section we discuss in detail the boundary worldsheet action of the brane/antibrane system, and set our conventions.

\subsection{Superspace action on the disk}
As a starting point, one considers the worldsheet action for coincident D1-brane and  anti-D1-brane wrapped around a circle in a compactified direction $Y$, T-dual 
to the system of interest. We set everywhere in the following $\alpha'=1$.

The $\mathcal{N}=(1,1)$ superspace action on the disk was written in~\cite{Hori:2000ic,Kraus:2000nj,Takayanagi:2000rz}, including the 
coupling to background gauge and tachyon fields. In the present context one considers non-trivial Wilson Lines along the circle, T-dual to the brane positions $x_1$ and $x_2$ 
along $X$, the T-dual of $Y$. They naturally appear in the form $x^{(\pm)} = x_1 \pm x_2$.  

Setting aside the 'spectator' dimensions, one considers a pair of $\mathcal{N}=(1,1)$ superfields on the disk, one time-like ($\mathbb{X}_0$) and the other compactified 
on a circle ($\mathbb Y$), with $e.g.$ $\mathbb X_0 = X_0 + \tfrac{i}{\sqrt{2}}( \theta \psi_0 +  \bar \theta \bar{\psi}_0) + \theta \bar \theta F_0$. The superspace coordinates 
are denoted as $\hat z=(z,\theta,\bar \theta)$.

At the boundary of the disk, the Grassmann coordinates satisfy the  boundary condition $\theta=\pm \bar \theta$.  The algebra of the Chan-Patton factors 
for the brane-antibrane system is conveniently implemented by the canonical quantization of boundary fermions~\cite{Marcus:1986cm}, see below. These boundary fermions are 
the bottom components of {\it Fermi superfields} of the boundary $\mathcal{N}=1$ superspace. For the brane-antibrane system one needs a complex superfield
\begin{equation}
\Gam^\pm = \eta^\pm + \theta F^\pm\, .
\end{equation}
with $\Gam^- = (\Gam^+)^*$.

Then the worldsheet action on the disk\footnote{Our convention is that any amplitude is computed with $e^{-S}$.}, including the tachyon background as well as Wilson lines around 
the circle, reads:
\begin{multline}
S_{\textsc{BCFT}}(\lambda^+,\lambda^-) = \frac{1}{2 \pi} \int_{D^2} \di^2 z\,  \di^2 \theta \; \parent{-D {\mathbb X}^0 \bar D {\mathbb X}^0 + 
D {\mathbb Y} \bar D {\mathbb Y}} +  i \oint_{S^1} \di u \,\di \theta \, \frac{x^{(+)}}{4\pi} D_u {\mathbb Y}  \\  
-  \oint_{S^1} \di u \, \di \theta \parent{\Gam^+ \parent{D_u + i \frac{x^{(-)}}{2\pi} D_u{\mathbb Y} } 
\Gam^- -  \Gam^+ {\mathbb T}^+ -  \Gam^- {\mathbb T}^-}\, ,
\label{eq:action_begin}
\end{multline}
with the  measure $\di^2 \theta = \di \theta\, \di \bar \theta$, the superspace holomorphic derivative $D = \partial_\theta + \theta \partial$ and the superspace boundary 
derivative $D_u = \partial_{\theta} + \theta \partial_u$, with the boundary coordinate $u$ on $S^1$.\footnote{ 
The boundary current superfield $D_u{\mathbb Y}$ is defined to be the boundary super-derivative of ${\mathbb Y}$ first taken to the boundary (where $\mathbb{Y}$ has 
Neumann boundary conditions). }

We consider simple rolling tachyon profiles of the form: 
\eqali{
{\mathbb T}^\pm  = \frac{\lambda^\pm}{2\pi}  e^{\omega {\mathbb X^0}}   \label{eq:tachyon}\, ,
}
with $0< \omega \leqslant \nicefrac{1}{\sqrt{2}}$. In order to get a real action, one chooses $(\lambda^+)^\ast = \lambda^-$.  These are actually the tachyons that we are 
expecting to be solutions of the spacetime effective action. It is understood in this expression that the superfield $\mathbb X$ is taken on the (super)boundary of the disk.

The space-time gauge field $A^{(-)}= -\tfrac{x^{(-)}}{4\pi} \, \text{d}y$ being locally pure gauge, its minimal coupling  to the Fermi superfields can be absorbed by a 'gauge' 
transformation.\footnote{This is a slight abuse of language, as this is {\it not} a gauge symmetry from the worldsheet perspective.} One has 
to be careful with this transformation if ${\mathbb Y}$-dependent insertions appear in the path-integral; a prescription must be chosen (see below). 
\eqali{
\Gam^\pm \to \Gam^\pm e^{\pm i \frac{x^{(-)}}{2\pi} {\mathbb Y}} \, .
}
After this field redefinition, the boundary Fermi superfields are free, with the  propagator on the real axis: 
\eqali{ \label{eq:gam_gam}
\corr{\Gam^+(\hat z)\Gam^-(\hat w)} = \,\hat \epsilon(\hat z-\hat w) = \,\epsilon(z-w)-2 \, \theta_z\theta_w \delta(z-w)  \, ,
}
with the sign function $\epsilon(z) = \Theta(z)-\Theta(-z)$. This implies that $\Delta(\Gam^\pm)=0$, $i.e.$ vanishing conformal dimension.

In terms of these new variables the worldsheet action~\refe{eq:action_begin} reads: 
\begin{multline}
S_{BCFT}(\lambda^+,\lambda^-) = \frac{1}{2 \pi} \int_{D^2} \di^2 z \, \di^2 \theta \; 
\parent{-D {\mathbb X}^0 \bar D {\mathbb X}^0 + D {\mathbb Y} \bar D {\mathbb Y}} +  i \oint_{S^1} \di u \, \di\theta  \frac{x^{(+)}}{4\pi} D_u {\mathbb Y}   \\ 
 -  \oint_{S^1} \di u\, \di\theta \, \parent{\Gam^+ D\Gam^- - \Gam^+ {\mathbb T}^+ -  \Gam^- {\mathbb T}^-} \label{eq:action}
\end{multline}
where the tachyon fields have now the expression: 
\eqali{\label{eq:tachprofiledual}
{\mathbb T}^\pm = \frac{\lambda^\pm}{2\pi} e^{\pm i \frac{x^{(-)}}{2\pi} {\mathbb Y} + \omega {\mathbb X}^0}\, .
}
Conformal invariance of the action at leading order imposes then : 
\eqali{
\omega^2 + \parent{\frac{x^{(-)}}{2 \pi}}^2 = \frac{1}{2}\, .
}
This is the standard mass-shell condition of an open string tachyon with $U(1)\times U(1)$ Wilson lines turned on.

The world-sheet action that describes a system of \emph{separated} brane and anti-brane is obtained from the previous one  
by a T-duality along $y$. In the bulk, the superfield $\mathbb Y$ is traded for the superfield $\mathbb X$ that has Dirichlet boundary conditions. Renaming 
$\mathbb Y$ as  $\tilde{\mathbb X}$, the tachyon interaction of interest reads 
\eqali{\label{tachprofile}
{\mathbb T}^\pm = \frac{\lambda^\pm}{2\pi} e^{\pm i \frac{x^{(-)}}{2\pi} \tilde {\mathbb X} + \omega {\mathbb X}^0}\, .
}

Action~\refe{eq:action} will be our starting point. In the free theory, one  has  two different boundary conditions on the disk boundary, 
related to the  distinct positions of the branes : $X = x^{(1)}$ or $Y=x^{(2)}$. We introduce the notations
\eqali{
& x^{(-)} = x^{(1)} - x^{(2)} = 2\pi r \nonumber \\
& x^{(+)} = x^{(1)}+x^{(2)}  = 2 x_{cm} \, ,
}
where on the first line $r$ is such that $\omega^2 + r^2 = 1/2$. On the second line, $x_{cm}$ is simply the center of mass coordinate of the system.

\subsection{Action in components, quantization of the Fermi superfields}
\label{subsec:compaction}
Starting from the action~\refe{eq:action},  renaming $\mathbb Y$ as  $\tilde{\mathbb X}$, and integrating over 
the fermionic coordinates one gets the action: 
\begin{multline}
S_{\textsc{BCFT}}(\lambda^+,\lambda^-) = \frac{1}{2 \pi} \int_{D^2} \di^2 z \; \parent{-\partial X^0 \bar \partial X^0 + 
\partial X \bar \partial X} +  i \oint_{S^1} \di u \, \frac{x^{(+)}}{4\pi} \partial_u \tilde X \\  
+  \oint_{S^1} \di u \parent{\eta^+ \partial_u  \eta^- - 
\frac{\lambda^+}{2\pi} ~ \eta^+ \psi^+ T^+ - \frac{\lambda^-}{2\pi} ~ \eta^- \psi^- T^- } \\ 
- \oint_{S^1} \di u \parent{F^+ F^- - F^+ ~ T^+  - 
F^- ~ T^-} \, ,
\label{eq:action_comp}
\end{multline}
with:
\eqali{\label{eq:def_tachyon}
\psi^\pm &= \pm i r \sqrt 2 \tilde \psi^x + \omega \sqrt 2 \psi^0 \nonumber \\
T^\pm &= e^{\pm i r \tilde X + \omega X^0}\, .
}
Auxiliary fields $F^\pm$ are then integrated to give: 
\begin{multline}
S_{\textsc{BCFT}}(\lambda^+,\lambda^-) = \frac{1}{2 \pi} \int_{D^2} \di^2 z \; \parent{-\partial X^0 \bar \partial X^0 + 
\partial X \bar \partial X} +  i \oint_{S^1} \di u \, \frac{x^{(+)}}{4\pi} \partial_u \tilde X \\  
+  \oint_{S^1} \di u \parent{\eta^+ \partial_u \eta^- - \frac{\lambda^+}{2\pi} \eta^+ \psi^+ T^+ - \frac{\lambda^-}{2\pi} 
\eta^- \psi^- T^- + \varepsilon^{1-4r^2} \frac{\lambda^+ \lambda^-}{4\pi^2} T^+ T^- }  \, .
\label{eq:action_int}
\end{multline}
A {\it contact term} at the end of the second line shows up, with a UV cutoff $\varepsilon$. This term, that does 
not follow from the equations of motion contributes nevertheless to correlation functions when $1/2<|r|<1/\sqrt{2}$. 
Its role will be discussed in  section~\ref{sec:contact_term}.

Finally, as the center-of-mass perturbation completely factorizes and commutes with any operators in~\refe{eq:action_int},  one can set $x^{(+)}=0$ without loss of generality.

Upon quantizing canonically the boundary fermions $\eta^\pm$, one recovers the Chan-Patton algebra corresponding to the brane-antibrane 
system~\cite{Takayanagi:2000rz}. It leads to the following identifications:
\eqali{
&\eta^+ \Leftrightarrow \sigma^+ = \frac{\sigma^1 + i\sigma^2}{2} \nonumber \\
&\eta^- \Leftrightarrow \sigma^- = \frac{\sigma^1 - i\sigma^2}{2} \nonumber \\
&\eta^+\eta^-(z) \Leftrightarrow \frac{\comm{\sigma^+}{\sigma^-}}{2} = \frac{\sigma^3}{2}
}
where now the prescription for the path integral is $Z=\text{Tr}\, \int \mathcal{D}X^i \mathcal{D}\psi^i P\, e^{-S[X^i,\psi^i]}$, 
which includes a {\it path ordering} for the operator insertions and a trace over the CP factors. In this context 
the tachyon becomes a {\it boundary changing operator}; when inserted on the boundary of the disk, 
it interpolates between the two distinct boundary conditions corresponding to the brane and to the anti-brane. 

The worldsheet action on the disk takes finally the form 
\eqali{\label{eq:wscompaction}
S = S_{bulk} - \oint_{S^1} \di u \parent{\frac{\lambda^+}{2\pi} \sigma^+ \otimes \psi^+ e^{i r \tilde X + \omega X^0} + \frac{\lambda^-}{2\pi} \sigma^- 
\otimes \psi^- e^{-i r \tilde X+ \omega X^0} - \frac{\lambda^+ \lambda^-}{4\pi^2} \varepsilon^{1-4r^2} e^{2 \omega X^0}}\, .
}

\section{Perturbative integrals and contact terms}
\label{sec:contactterms}
In this section we discuss in more detail the contact term, quadratic in the tachyon field, that appear in the action~(\ref{eq:action}) after integrating out the  
auxiliary fields from the Fermi superfields $\Gam^{\pm}$, and quantizing their fermionic components. As was discussed long ago by Green and 
Seiberg~\cite{Green:1987qu} for closed string correlation functions,  contact terms, dictated by worldsheet supersymmetry, can cancel unphysical divergences in 
correlation functions.  We shall see below that it indeed cancels the short-distance singularity when two tachyons perturbations collide in the perturbative expansion.

\subsection{Free field correlators}
In order to fix the conventions, we use the following Green functions on the upper half-plane $H^+$ for a free-field $X$ with Dirichlet boundary conditions,  and its 
T-dual field $\tilde X$:
\eqali{
\corr{X(z_1)X(z_2)} &= -\frac{\eta_{xx}}{2} \ln |z_{12}|^2 + \frac{\eta_{xx}}{2}\ln |z_{1\bar 2}|^2  \nonumber \\ 
\corr{\tilde X(z_1)\tilde X(z_2)} &= -\frac{\eta_{xx}}{2}\ln |z_{12}|^2 - \frac{\eta_{xx}}{2}\ln |z_{1\bar 2}|^2  \nonumber \\
\corr{X(z_1)\tilde X(z_2)} &= -\frac{\eta_{xx}}{2}\ln \frac{z_{12}}{z_{\bar 1 \bar 2}} - \frac{\eta_{xx}}{2}\ln \frac{z_{1\bar 2}}{z_{\bar 1 2}}  
}
with \eg $z_{12} = z_1  - z_2$ and $z_{1 \bar 2} = z_1 - \bar z_2$. Finally, the two-point function for fermions with Dirichlet b.c. read:
\begin{subequations}
\begin{align}
\corr{\psi^x(z_1)\psi^x(z_2)} &= \frac{\eta^{xx}}{z_1-z_2}\\
\corr{\bar \psi^x(\bar z_1) \bar \psi^x(\bar z_2)} &= \frac{\eta^{xx}}{\bar z_1-\bar z_2}\\
\corr{\psi^x(z_1)\bar \psi^x(\bar z_2)} &= - \frac{\zeta \eta^{xx}}{z_1-\bar z_2} \label{mixedfer}
\end{align}
\end{subequations}
where $\zeta=\pm 1$ corresponds to the spin structure. It corresponds to the boundary conditions for the supercurrent 
$G (z)- \zeta \bar{G} (\bar z)|_{z=\bar z}=0$. For the Virasoro superfield $\mathbb G = G + \theta T$,  this is naturally 
associated with the superspace boundary $(z,\theta)=(\bar z,\zeta \bar \theta)$.  With Neumann b.c., eq.~(\ref{mixedfer}) gets a minus sign on the RHS.

Finally, the boundary Green function for a superfield $\mathbb{X}$ with Neumann boundary conditions reads:
\begin{equation}
\langle \mathbb{X} (\hat z_1) \mathbb{X} (\hat z_2)\rangle_{\Im z=0,\theta=\zeta \bar \theta} = -2\eta_{xx} \ln \hat{z}_{12} =  -2\eta_{xx} \ln (z_{12} - \theta_1 \theta_2)
\end{equation}
while it vanishes with Dirichlet b.c..

\subsection{Contact term in the worldsheet action} \label{sec:dist_pert}
As has been explicited in section~\refe{subsec:compaction}, upon integrating out the auxiliary fields $F^\pm$ that appear in the Fermi multiplets $\Gam^\pm$, one obtains a contact 
term for the tachyon in the worldsheet action.

The auxiliary field has the two-point function \mbox{$\langle F^+ (u) F^- (v) \rangle =2\delta (u-v)$}. It is regularized at short distances according to: 
\begin{equation}
\corr{F^+(t)F^-(s)}=2\delta(t-s) \to \delta(\module{t-s}-\varepsilon) 
\end{equation}
It was shown in~\cite{Gaberdiel:2009hk} that this point-splitting regularization that we use preserves worldsheet supersymmetry (unless one consider 
bulk-boundary correlators for which more care is needed). 

Then the contact term is given by the following non-local  interaction on the disk (with $u=e^{it}$, $v=e^{is}$): 
\begin{multline}\label{actwithcontact}
\frac{1}{2}\int_0^{2 \pi} \di t \int_0^{2\pi} \di s \, \delta(\module{t-s}-\varepsilon) \, \bnormal{e^{i r \tilde X +\omega X^0}(u)}\, \bnormal{e^{-i r \tilde X +\omega X^0}(v)} \\
\displaybreak[2]
= \frac{1}{2} \int_0^{2 \pi} \di t \int_0^{2\pi} \di s \, \delta(\module{t-s}-\varepsilon)\module{u-v}^{2(\omega^2-r^2)} 
\bnormal{e^{i r \tilde X +\omega X^0}(u)\,e^{-i r \tilde X +\omega X^0}(v)}\\ 
= \frac{1}{2}  \parent{2\sin \frac{\varepsilon}{2} }^{1-4 r^2} \int_0^{2\pi}\!\! \!\! \di s ~  \Bigg( 
\! \bnormal{e^{i r \tilde X +\omega X^0}(v+\varepsilon)\,e^{-i r \tilde X +\omega X^0}(v)} 
 + \bnormal{e^{i r \tilde X +\omega X^0}(v)\,e^{-i r \tilde X +\omega X^0}(v+\varepsilon)} 
\Bigg) \\
 \stackrel{\varepsilon \to 0}{\sim} \varepsilon^{1-4r^2}  \oint_{S^1} \di u ~ \bnormal{e^{2\omega X^0}(u)}\, .\end{multline}
By $\bnormal{\star}$ we denote the boundary normal ordering (see $e.g.$~\cite{Polchinski:1998rq}).\footnote{We added a $\nicefrac{1}{2}$ normalization such that to 
take account for the factor 2 coming from the trace over the CP factor, since the contact term is multiplied by the identity matrix.} This treatment of the contact term may 
seem a bit {\it ad hoc}, however we will find in the next section that  the term~(\ref{actwithcontact}) appears naturally 
when one considers the renormalisation of the worldsheet action, justifying {\it a posteriori} this presentation.

We will use in a next section the contact term on the upper half plane. It is similarly written as: 
\begin{multline} \label{eq:cont_upper}
\frac{\varepsilon^{1-4r^2} }{2} \int_{-\infty}^{+\infty}\!\! \di v \, 
\Bigg( \bnormal{e^{i r \tilde X +\omega X^0}(v+\varepsilon)\,e^{-i r \tilde X +\omega X^0}(v)} + 
\bnormal{e^{i r \tilde X +\omega X^0}(v)\,e^{-i r \tilde X +\omega X^0}(v+\varepsilon)} \Bigg) \\
 \stackrel{\varepsilon \to 0}{\sim} \varepsilon^{1-4r^2}  \oint_{\mathbb R} \di v ~ \bnormal{e^{2\omega X^0}(v)}
\end{multline}
 In order to compute all the counterterms generated from this contact term one will need to work with its complete non-local expression, though the dominant term,
here the only divergent one, in its Taylor expansion (in terms of local operators) is sufficient to compute most of them. 
Indeed, it is found that working directly with the dominant term, a local operator, seems to be equivalent to working with the complete non-local contact term. 
It may be explained by the fact that after Taylor expansion of $T^\pm (x+\epsilon)$ and commutation of the sum and the integral, 
all other terms in the series of integrated local operators vanish as $\epsilon$ goes to zero. One may object that we are forgetting sub-dominant terms, but, 
as the UV cut-off is an artifact signaling our lack of ability to manipulate infinite quantities; it is to be understood as being strictly equal to zero, 
from the very beginning. From this point of view, we expect that only the divergent 
terms in~\refe{eq:cont_upper} do contribute. Then it should be equivalent to use either the dominant (local) term or the complete (non-local) contact term. 
This statement seems to be confirmed numerically in the fourth order computations of section~\ref{sec:secbeta}.

As one can see, in the limit $\varepsilon  \to 0$ when one takes the UV cut-off to infinity, the contact term vanishes when $|r|<1/2$. Therefore, the 
results of the computations made in~\cite{Sen:2003tm}, where the contact term was not taken into account, remain unchanged.\footnote{As a side remark, for 
the rolling tachyon on a non-BPS D-brane, it was already noticed in~\cite{Fotopoulos:2003yt} that the contact terms, that were absent in the original 
computation of the partition function performed in~\cite{Larsen:2002wc}, did not contribute to the final result.} It can be seen also by 
working directly with the $\mathcal{N}=1$ boundary superspace amplitudes; the contact terms contributions from the $\Gam^\pm$ 
correlators vanish for $|r|<1/2$.

However, the contact term diverges when $|r|>1/2$. This contact term may  ensure that the amplitudes do not diverge for $|r|>1/2$. 
The divergence associated with the contact term, that arises from the fusion of two tachyon vertices, correspond to the unphysical integrated vertex operator
\begin{equation}\label{eq:resonanceterm}
\int \di u  \int \di \theta \, \theta \, \bnormal{ e^{2\omega \mathbb{X}_0 (u)}} = 
\int \di u \, \bnormal{ e^{2\omega X_0 (u)}} \, ,
\end{equation}
that is not supersymmetric. Hence, as in~\cite{Green:1987qu}, one can understand the contact term as necessary  to preserve worldsheet superconformal 
invariance on the boundary, when $|r|>1/2$. In other words, divergences corresponding to integrated operators of the form~(\ref{eq:resonanceterm}) cannot occur for a 
consistent, hence super-BRST invariant, superstring worldsheet theory. We will discuss below higher-order divergences, coming from the fusion of more than two operators, 
for which the analysis is more involved.

\subsection{Boundary one-point function} \label{sec:contact_term}
In order to illustrate more precisely the role of the contact term, we compute the  one-point function on the disk for a tachyon boundary vertex operator. 
This one-point function does not have to vanish because of the rolling tachyon background, and contains potentially a divergence at first order, 
when the inserted tachyon vertex collides with the integrated tachyon coming from the perturbative expansion. We will find 
that the contact term cancels the two-tachyon divergence for all values of $r$ in the range  $1/2<|r| \leqslant 1/\sqrt 2$.

At first order in the couplings $\lambda^\pm$, the one-point function for one of the boundary tachyon vertex operators is given by 
the integrated correlator 
\begin{multline}
 \tr \corr{\sigma^\pm \otimes \psi^\pm 
e^{\pm i r \tilde X + \omega X^0 }(e^{it_1})}  \\ \sim \frac{\lambda^\mp}{2 \pi} \tr \sigma^\pm \sigma^\mp \int_{0}^{t_1} 
\! \di t_2 \corr{\psi^\pm e^{\pm i r \tilde X + \omega X^0 }(e^{it_1})\,  \psi^\mp e^{\mp i r \tilde X + \omega X^0 }(e^{it_2})}_0  \\  
 + \frac{\lambda^\mp}{2 \pi} \tr \sigma^\mp \sigma^\pm \int_{t_1}^{2 \pi}\! \di t_2 \corr{\psi^\mp e^{\mp i r \tilde X + \omega X^0 }(e^{it_2}) \, 
\psi^\pm e^{\pm i r \tilde X + \omega X^0 }(e^{it_1})}_0  \\ 
   \sim   \frac{\lambda^\mp}{2 \pi} \parent{1-4r^2} \int_{t_1}^{t_1 + 2 \pi}\! \di t_2 \module{2 \sin \frac{t_1-t_2}{2}}^{2\omega^2 - 2r^2 -1} 
  \int_{-\infty}^{+\infty} \di x^0 \, e^{2 \omega x^0}\, .
\end{multline}
The integration over $t_2$ is not defined for $|r|>1/2$, nevertheless the result
\eqali{
 \tr \corr{\sigma^\pm \otimes \psi^\pm 
e^{\pm i r \tilde X + \omega X^0 }(e^{it_1})} 
\sim  \frac{\lambda^\mp}{2 \pi} \parent{1-4r^2} 2^{1 - 4 r^2} \sqrt \pi \frac{\Gamma(\frac{1}{2}-2r^2)}{\Gamma(1-2r^2)}
  \int_{-\infty}^{+\infty} \di x^0\,  e^{2 \sqrt{\frac{1}{2}-r^2} x^0}}
is analytic for any $r\in[0,1/\sqrt 2]$

In order to show how the divergence for $|r|>1/2$ is canceled, we can compute directly this quantity in superspace, using the 
Fermi multiplets $\Gam^\pm$. Letting aside for a moment the zero-mode integral over $x_0$, one considers the superspace integral 
\begin{multline}
\int \di\theta_1 \corr{\Gam^\pm e^{\pm i r \tilde {\mathbb X} + \omega {\mathbb X^0} }(\hat z_1)} \\ \sim  - \frac{\lambda^\mp}{2 \pi}
\int \di\theta_1 \di\theta_2 \int \di t_2 \, \epsilon (\hat z_1-\hat z_2) \corr{e^{\pm i r \tilde {\mathbb X} + \omega {\mathbb X^0} }(\hat z_1) 
e^{\mp i r \tilde {\mathbb X} + \omega {\mathbb X^0} }(\hat z_2)}_0 \\ 
  \sim  -\frac{\lambda^\mp}{2 \pi} e^{2\omega x^0}\int \di \theta_1 \di \theta_2 \int \di t_2  \left[ \epsilon(t_1-t_2) - 2 \theta_1\theta_2 \delta(t_1-t_2)\right] \times 
\hphantom{aaaaaaAA}
\\ 
\hphantom{aaaaaaaaaaaaaaaaaaaaaAA}\times
\parent{\module{2\sin\frac{t_1-t_2}{2}}^{1-4r^2}\!\!\!\!\!\! - \theta_1\theta_2 (1-4r^2)\epsilon(t_1-t_2) \module{2\sin\frac{t_1-t_2}{2}}^{-4r^2}}  \\ 
  \sim  -\frac{\lambda^\mp}{2 \pi} e^{2 \omega x^0} \int \di t_2 \croch{(1-4r^2) \module{2\sin\frac{t_1-t_2}{2}}^{-4r^2} + 
2 \delta(t_1-t_2)\module{2\sin\frac{t_1-t_2}{2}}^{1-4r^2}} \, .
\end{multline}

Now we introduce a point splitting regularization, asking that $\module{t_1-t_2}>\varepsilon$. As we wish to keep the contact term in 
the computation, it is natural to include this point splitting in the  $\Theta$ and $\delta$ distributions that appear in the above integral, as: 
\eqali{
&\Theta(\module{t_1-t_2}-\varepsilon) = \Theta(t_1-t_2-\varepsilon)+\Theta(t_2-t_1-\varepsilon) \nonumber \\ 
&\delta(\module{t_1-t_2}-\varepsilon) = \delta(t_1-t_2-\varepsilon)+\delta(t_2-t_1-\varepsilon) \, .
} 
In other words, we 'spread' the contact term at the boundary of the interval $\module{t_1-t_2}<\varepsilon$. Then 
the contribution to the one point-function becomes:
\eqali{
 &-\frac{\lambda^\mp}{2 \pi}  \int \di t_2 \croch{(1-4r^2)\Theta(\module{t_1-t_2}-\varepsilon) \module{2\sin\frac{t_1-t_2}{2}}^{-4r^2} + 
\delta(\module{t_1-t_2} - \varepsilon)\module{2\sin\frac{t_1-t_2}{2}}^{1-4r^2}} \nonumber \\ 
& = -\frac{\lambda^\mp}{2 \pi} (1-4r^2) \int_{t_1-2\pi+\varepsilon}^{t_1-\varepsilon} \di t_2 \module{2\sin\frac{t_1-t_2}{2}}^{-4r^2} -  
2 \frac{\lambda^\mp}{2 \pi} \module{2\sin\frac{\varepsilon}{2}}^{1-4r^2} \nonumber \\
& \sim -\frac{\lambda^\mp}{2 \pi} \parent{1-4r^2}   2^{1 - 4 r^2} \sqrt{\pi}\frac{\Gamma(\frac{1}{2}-2r^2)}{\Gamma(1-2r^2)} + 
2\frac{\lambda^\mp}{2 \pi} \varepsilon^{1-4r^2} -  2\frac{\lambda^\mp}{2 \pi} \module{2\sin\frac{\varepsilon}{2}}^{1-4r^2} \, ,\label{eq:onepfinteg}
} 
where two first terms in the last line come from the expansion of the following function: 
\eqali{
\parent{1-4r^2} 2^{2-4r^2} \cos\frac{\varepsilon}{2} ~ {}_2 F_1 \croch{\frac{1}{2},\frac{1+4r^2}{2},\frac{3}{2},\cos^2 \frac{\varepsilon}{2}} \, .
}
The second term of eq.~(\ref{eq:onepfinteg})  is the only divergent one if $4r^2 > 1$. It simplifies to  
\eqali{
-\frac{\lambda^\mp}{2 \pi} \parent{1-4r^2} 2^{1 - 4 r^2} \sqrt \pi \frac{\Gamma(\frac{1}{2}-2r^2)}{\Gamma(1-2r^2)} + 
2 \frac{\lambda^\mp}{2 \pi} \varepsilon^{1-4r^2} - 2 \frac{\lambda^\mp}{2 \pi} \varepsilon^{1-4r^2} \, .
}
Divergences compensate correctly, so that we eventually have at first order: 
\eqali{
\tr \corr{\parent{\sigma^\pm \otimes \psi^\pm - F^\pm} e^{\pm i r \tilde X + \omega X^0 }(z_1)} & \sim  -  
\lambda^\mp  \frac{\Gamma(2-4r^2)}{\Gamma^2(1-2r^2)}  \int_{-\infty}^{+\infty} dx^0 e^{2 \sqrt{\frac{1}{2}-r^2} x^0} \, .
}
This quantity is UV-finite, but has a IR divergence due to the zero-mode integral. This divergence, that appears when $x^0 \to \infty$, 
simply signals the breakdown of perturbation theory in $\lambda^\pm$.  Note that for the homogeneous rolling tachyon on a non-BPS brane, for which the all orders computation is 
doable, summing up the the whole perturbative expansion  gives a finite zero-mode integral.\footnote{If we Wick-rotate the theory to an Euclidean target space,  for which 
perturbation theory is well-defined, the zero-mode integration gives $\delta(2\omega)$ which is zero for any value of $|r|<1/\sqrt 2$.}

\section{Computation of beta-functions}
\label{sec:secbeta}
In this section, we argue that the theory defined in~\refe{eq:action} is exactly conformal, with the rolling tachyon profile~(\ref{tachprofile}), for 
any value of $|r|$ below $r_c = 1/\sqrt{2}$. This will imply that for the spacetime effective action of the brane-antibrane system there exists a 
'half S-brane' rolling tachyon solution at fixed separation of the equations of motion. This is an important point since the effective action proposed in~\cite{Garousi:2004rd} 
did not admit solution at fixed distance~; in fact, in this action, for non-vanishing tachyon the distance field has an attractive potential 
towards the origin.

Our motivation for looking closely at this issue was in part due to the results of Bagchi and Sen~\cite{Bagchi:2008et}. They found that the boundary 
deformation corresponding to the tachyon~(\ref{tachprofile}) was only marginal in the range $0\leqslant |r| < r_c/\sqrt{2}$. For $r_c/\sqrt{2}\leqslant |r|<r_c$ 
it was found that for an infinite but countable set of distances the theory was not conformal. This is puzzling as we expect 
that everything goes smoothly up to the critical separation $r_c$. 

At the end of the day, the basic difference between those two approaches is the contact term,  however the latter is not responsible for restoring marginality, since 
it cannot cancel the logarithmic divergences that could spoil conformal invariance as we shall see; rather, the actual computation of the possible conformal symmetry-violating 
terms in the path integral gives zero thanks to the different contributions that cancel among themselves at a given order. Nevertheless, the contact term is able, 
as expected, to cancel the power-like two-tachyon divergences in the perturbative integrals.

The cleanest way to show that the action~(\ref{eq:action}), with the rolling tachyon perturbation~(\ref{eq:tachprofiledual}) is a boundary CFT 
is to compute the boundary $\beta$-functions for all the boundary couplings involved. On top of the coupling constants $\lambda^\pm$ for the rolling tachyon 
perturbations, one needs to introduce in the computation a  perturbation corresponding to the separation-changing boundary 
operator $\sigma^3 \otimes i\partial_u X$.\footnote{To be exact we will have 
to add it in superspace as $\Gam^+\Gam^- D{\mathbb X}$}  The brane-antibrane separation is classically 
fixed at some value $r$, but still in the quantum  theory one has to check that the corresponding beta-function vanishes for any $r$, 
in other words that it is not 'sourced' by terms in $\lambda^{\pm}$. On top of this, more operators need to be considered in the 
analysis as $|r|$ increases.

\subsection{Generalities about boundary beta-functions}
In order to compute the beta-functions for their boundary couplings, we follow mostly the clear presentation of~\cite{Gaberdiel:2008fn}.

One considers a conformal field theory on the upper half-plane $H^+ =\sing{z,\im~ z \geq 0}$  perturbed by boundary operators that can be marginal or relevant. 
The action of the theory is defined to be 
\begin{equation}
\label{eq:renormaction}
S (\lambda^\mu) = S_{bulk} + \sum_\mu \ell^{-y_\mu} \lambda^\mu \int \di x\,  \phi_\mu (x) +S_{ct}\, ,
\end{equation} 
in terms of the renormalized dimensionless couplings $\{\lambda^\mu\}$ and the anomalous dimensions $y_\mu=1-h_\mu$. The renormalization scale 
is denoted by $\ell$. The last term $S_{ct}$ stands for  boundary counterterms whenever they are necessary. The boundary fields 
$\phi_\mu$ are normalized as\footnote{The Zamolodchikov correlators are defined as  
$(\phi_a(\infty)| \phi_b(z_b))  = \lim_{z \to \infty} z^{2 h_a}\bar z^{2 \bar h_a} \corr{\phi_a(z)\phi_b(z_b)}$.} 
\eqali{
\zamcorr{\phi^*_\mu (\infty)}{\phi_\mu(0)} = 1
} 
with $\phi^*_\mu$ the conjugate field to $\phi_\mu$.\footnote{In the case of theories with several boundary conditions, 
one has to trace over the  Chan-Patton factors, which would be here included inside the 
fields, $e.g.$ as $\tr\zamcorr{\phi^{*}_\mu (\infty)}{\phi_\mu(0)} = 1$. Considering deformations by boundary-changing operators,
the CP factors induce selection rules.}

At second order in perturbation theory, one encounters the integral (which lies inside a correlator with arbitrary other insertions):\footnote{
We will use the convention that operators (with CP factors) are ordered from right to left with increasing boundary parameter; this is the opposite 
convention than in~\cite{Gaberdiel:2008fn}.}
\begin{equation}
\label{eq:intbeta}
\frac{1}{2} \sum_{\mu,\nu} \ell^{h_\mu-h_\nu-2} \int \di x_1 \int \di x_2 \, 
\phi_\mu (x_1) \phi_\nu (x_2) \Theta (|x_1-x_2|-\varepsilon) \Theta (L-|x_1-x_2|)\, 
\end{equation}
This integral has been regularized by point-splitting with a UV cutoff $\varepsilon$ , and with and an IR cutoff $L$. 
In order to compute the integral one can use the boundary OPE
\begin{equation}
\phi_\mu (x_1) \phi_\nu (x_2) = \sum_{\rho} \frac{D_{\mu \nu}^{\rho}}{(x_1-x_2)^{h_\mu+h_\nu-h_\rho}} \phi_\rho (x_2) + \cdots \qquad x_1>x_2  \, .
\end{equation}
In this case,~\refe{eq:intbeta} is rewritten as:
\begin{equation}
\sum_{\mu < \nu} \ell^{h_\mu-h_\nu-2} \parent{\oint_{y+\epsilon}^{y+L} \di x \oint \di y ~ \phi_\mu(x) \phi_\nu(y) + \oint^{y+L}_{y+\epsilon} 
\di x \oint \di y ~ \phi_\nu(x) \phi_\mu(y)}
\end{equation}

\subsubsection*{Minimal substraction scheme}
In this scheme, we aim to isolate the divergences that occur in the integral~(\ref{eq:intbeta}) when  two perturbations collide. 
One has to consider separately two cases. The subset of boundary fields $\{ \phi_\rho\}$ such that $y_\mu+y_\nu-y_\rho <0$ (which are all relevant), 
gives a divergent contribution to the action~(\ref{eq:renormaction}) of the form (after removing the IR cutoff) 
\begin{equation}
S_d = \frac{1}{2} \sum_{\mu,\nu,\rho} \frac{D_{\mu \nu}^{\rho}}{y_\mu+y_\nu-y_\rho} \varepsilon^{y_\mu+y_\nu-y_\rho}
\ell^{y_\mu+y_\nu} \lambda^\mu \lambda^\nu \int \di x \, \phi_\rho \, .
\end{equation}
In the minimal substraction scheme, this divergence is canceled by a similar counter term $S_{ct}=-S_d$. 

The subset of boundary fields  $\{ \phi_\tau \}$ such that $y_\mu+y_\nu-y_\tau =0$ gives  logarithmic divergences, or resonances 
(cutting the integration at the renormalization scale $\ell$):
\begin{equation}
S_d = \frac{1}{2} \sum_{\mu,\nu,\tau} D_{\mu \nu}^{\tau} \ln (\varepsilon/\ell)
\ell^{-y^\tau} \lambda^\mu \lambda^\nu \int \di x \, \phi_\tau
\end{equation}
This divergent piece is again canceled by an appropriate counterterm $S_{ct}=-S_d$. Now equating the bare couplings to the two corresponding contributions from 
the renormalized action~(\ref{eq:renormaction}), one gets the beta-function at second order
\begin{equation}
\label{eq:beta}
\beta^{\textsc{ms}}_\rho := \ell \frac{\di \mu_\rho}{\di \ell}= y_\rho \lambda^\rho - 
\!\!\! \sum_{\mu,\nu | y_\mu+y_\nu=y_\rho} \!\! D_{\mu \nu}^{\rho}  \lambda^\mu \lambda^\nu
\end{equation}
So  non-linear contributions at quadratic order occur only in the cases of resonances, if they exist.\footnote{Notice that, if the boundary perturbations in~(\ref{eq:intbeta}) 
are superficially marginal, the resonances correspond to the appearance of a marginal operator in the boundary OPE.} One can show that, in the minimal 
substraction scheme, this property holds to all orders in perturbation theory. 

Note that there is a sign difference between the above result and what appears in~\cite{Gaberdiel:2008fn}. This comes from their convention of using $e^{S}$ instead of $e^{-S}$ as we did. One could obtain the same definition by simply changing the sign of the couplings.

\subsubsection*{Wilsonian scheme}
In this scheme, we equate the renormalization scale $\ell$ with the UV scale $\varepsilon$, viewed as a fundamental high-energy scale. We demand that 
the renormalized theory does not depend on the UV cutoff scale, i.e. that $\varepsilon \partial_\varepsilon e^{-S_{bdy}} = 0$. Then 
 the renormalized boundary couplings depend on the UV scale $\varepsilon$ (as the regularized perturbative integrals do). At second order, the corresponding beta-functions read: 
\begin{equation}
\label{eq:betaw}
\beta^{\textsc{ws}}_\rho :=\varepsilon \partial_\varepsilon \mu_\rho= y_\rho \lambda^\rho - \sum_{\mu,\nu} \!\! D_{\mu \nu}^{\rho}  \lambda^\mu \lambda^\nu
\end{equation}
In contrast with the minimal substraction scheme, eq.~(\ref{eq:beta}), there is no restriction to 'resonant' boundary couplings 
in the sum giving the quadratic term of the beta-function~(\ref{eq:betaw}).\footnote{The linear term, as well as the resonant quadratic terms, that 
are common to both schemes, can be shown to be 'universal', $i.e.$ independent of the scheme chosen for the computations.}

We will see below that both schemes are useful in the study of the rolling tachyon perturbations, when it comes to understand the role of the contact terms.

\subsection{Beta-functions for the brane-antibrane system at second order}
Coming back to the brane-antibrane system, we consider the following worldsheet action on the upper half-plane, as a function of the boundary 
couplings. So now we take the boundary variable to be $u\in{\mathbb R}$. For convenience, we rescale the coupling according to $\lambda^\pm \to 2 \pi \lambda^\pm$.
\begin{equation}
S
= S_{bulk} - \int \di x \Big( \lambda^+ \sigma^+ \otimes \psi^+ e^{ir \tilde X + \omega X^0} + \lambda^- \sigma^- \otimes \psi^- e^{-ir \tilde X+ \omega X^0}
 - i \frac{\delta r}{2} \sigma^3 \otimes \partial_u \tilde X \Big)   \label{eq:act_pert}
\end{equation}  
We omitted for the moment the contact term, which will enter later on in the discussion.

\subsubsection*{Distance coupling}
Let us start by discussing the beta-function for the distance perturbation. According to the general discussion above,
one has
\begin{equation}
\label{eq:betadistance}
\beta_r = (1-h_r) \frac{\delta r}{2} - (D^r_{+ -}  + D^r_{- +}) \lambda^- \lambda^+ - D^r_{r +} \frac{\delta r}{2} \lambda^+ -  D^r_{r -} \frac{\delta r}{2} \lambda^-\ldots
\end{equation}
where the ellipsis here stands for higher order terms. The first term on the RHS vanishes because the conformal dimension of the 
distance perturbation is one. All the second order, all the structure constants for the three boundary operators under study appear, since, being all 
of conformal dimension one, they lead potentially to resonances. 

Without much work, we have that $D^r_{\pm \mp}=0$. The fusion of the tachyon vertex operators $T^\pm$ will never produce the current $\partial_u X$,  as t
he $e^{\omega X_0}$ factors just add up. The structure constants $D^r_{r \pm}$ also have to vanish, since 
the fusion of $T^\pm$ with the boundary current $i \sigma^3\partial_u \tilde X$ comes with the Chan-Patton factor 
$\sigma^\pm \sigma^3=\mp \sigma^\pm$ hence not $\sigma^3$. However, this product participates to the beta-function of $T^\pm$ as we will see.

At higher orders in perturbation theory, we would find a similar behavior. Namely, the fusion of any number tachyon vertices cannot produce the distance-changing 
operator, hence the beta-function for $\delta r$ does not get tachyon 'source terms' (which would be proportional to $(\lambda^+ \lambda^-)^n$ at order $2n$). In other words, 
the distance coupling does not run in the rolling tachyon background~(\ref{tachprofile}).

We can also be less specific and consider, instead of~(\ref{tachprofile}),  a more general tachyon profile of the form:
\begin{align}\label{eq:tachyons}
{\mathbb T}^\pm  =  \frac{1}{2\pi} e^{\pm i r \tilde {\mathbb X}} \parent{\lambda^\pm \, e^{ \omega {\mathbb X}^0}  + \xi^\pm \, e^{- \omega {\mathbb X}^0}}  \, ,
\end{align}
the hermiticity of the action imposing that $\lambda^- = \overline{\lambda^+}$ and $\xi^-=\overline{\xi^+}$. 

The conclusion can be different, as the structure constants $D^r_{\pm \mp}$ do not have to vanish by similar arguments. To this end, we use the 0-picture tachyons OPE:
\begin{align}
\sigma^+ \sigma^- \otimes T_{(0)}^+(z)T_{(0)}^-(w) & = \frac{1+\sigma^3}{4} \otimes \parent{\ldots + i r \, \frac{\partial_u \tilde X}{z-w} + \ldots} \nonumber \\ 
\sigma^- \sigma^+ \otimes T_{(0)}^-(z)T_{(0)}^+(w) & = \frac{1-\sigma^3}{4} \otimes \parent{\ldots - i r \, \frac{\partial_u \tilde X}{z-w} + \ldots}
\end{align}
where we only highlighted the interesting term. It is not difficult then to obtain the second-order beta-function for the distance coupling~:
\begin{equation}\label{generalbetadistance}
\beta_{r} = - \frac{\lambda^+ \xi^- + \lambda^- \xi^+}{4\pi^2} \, r = - \frac{1}{2\pi^2} \,\re \left(\lambda^+ \overline{\xi^+} \right) \, r 
\end{equation}
The beta-function~(\ref{generalbetadistance}) is scheme-independent, as the divergence is logarithmic. If one introduces a real parameter $\mu$, 
the most general solution of this equation is then:
\begin{equation}\label{generictachyonprofile}
{\mathbb T}^\pm  = \frac{\lambda^{\pm}}{2\pi}\, e^{\pm i r \tilde {\mathbb X}} \parent{ e^{\omega {\mathbb X}^0} \pm i \mu \, e^{-\omega {\mathbb X}^0}} \ , \quad \mu \in \mathbb{R}
\end{equation}
Notice that, so far, the marginality of this solution has been checked  only at second order. The fourth order (and higher) beta-functions would be non-trivial to compute, 
and marginality at this order is \emph{a priori}  not obvious. 

In any case, the physical meaning of the general solution~(\ref{generictachyonprofile}) is not clear. If, for instance, one chooses $\lambda^+$ real 
(which is always possible by a shift of $\tilde{X}$), it corresponds to a case where the real part of the tachyon condenses, while the imaginary part evolves in the 
opposite direction. There is no clear reason why the phase of the tachyon condensate {\it has} to change by $\pi/2$ during its evolution, as the constraints on the 
solution~(\ref{generictachyonprofile}) suggests. By symmetry arguments, the tachyon potential of the effective action should depend only on its square modulus. $i.e.$ of 
\begin{align}
\module{T(x^0)}^2 = \frac{\lambda^+\lambda^-}{4\pi^2}\left( e^{2 \omega x^0} + \, \mu^2 \, e^{-2 \omega x^0} \right)\, ,
\end{align}
which, for non-zero $\mu$,  goes through a minimal value $|T|^2 = \mu \frac{\lambda^+\lambda^-}{2\pi^2}$ at finite time.

If we consider instead the time-reversal-symmetric tachyon profile ('full S-brane') as in the case of the non-BPS brane:
\begin{equation}\label{fullSbrane}
\mathbb{T}^{\pm} = \frac{\lambda^\pm}{2\pi} e^{\pm i r \mathbb{X}} \cosh \omega \mathbb{X}_0\, ,
\end{equation}
the beta-function~(\ref{generalbetadistance}) indicates a RG running of the distance coupling, unless $r=0$. 
This result has far-reaching consequences. Unlike the case of coincident brane-antibrane or of a non-BPS brane, the effective action of the brane-antibrane system 
at finite distance  should be such that, while the 'half S-brane' rolling tachyon is allowed as a solution of its equations of motion, the 'full S-brane' should not.

\subsubsection*{Tachyon couplings at quadratic order}
We now compute the beta-functions for the tachyon couplings $\lambda^\pm$ at order 
$\lambda^+ \lambda^-$ for the 'half S-brane' profile.\footnote{Let us remark in passing that, by shifting the zero-mode of the time-like field 
$X_0\to X_0 + \alpha$, there is a common rescaling of the couplings $\lambda^\pm \to \lambda^\pm e^{\omega \alpha}$. This  is a common feature of Liouville-like 
theories. For this reason, the perturbative expansion in $\lambda^\pm$ does strictly make sense only in the Euclidean theory obtained by $X_0 \to i X_\textsc{e}$.}

The boundary OPEs to consider at quadratic order are the distance-tachyon OPE 
\begin{equation}\label{eq:disttachope}
-i\sigma^3 \otimes \partial_u \tilde{X} (x_1) \, \sigma^{\pm}\otimes \psi^{\pm} e^{\omega X_0 \pm ir\tilde{X} } (x_2)
\sim \frac{-2}{x_1-x_2} (\pm \sigma^\pm) \otimes (\pm r)  \psi^{\pm} e^{\omega X_0 \pm ir\tilde{X} } (x_2) + \cdots
\end{equation}
and the tachyon-tachyon OPE
\begin{multline}\label{eq:tachtachope}
\sigma^{+}\otimes \bnormal{\psi^{+}  e^{\omega X_0 + ir\tilde{X} } (x_1)}\, 
\sigma^{-}\otimes \bnormal{\psi^{-}  e^{\omega X_0 - ir\tilde{X} } (x_2)}\\
\sim  -  \Big( \begin{array}{cc} 1 & 0 \\ 0& 0 \end{array}\Big) (1-4r^2) \frac{1}{(x_1-x_2)^{4r^2}}\,  \bnormal{e^{2\omega X_0}(x_1)}   +\cdots 
\end{multline}
both for $x_1 > x_2$. The ellipsis stands for less singular terms.

\paragraph{\boldmath Beta-function for $|r|<1/2$.}
Whenever $|r|<1/2$  the OPE~(\ref{eq:tachtachope}) does not lead to singularities when integrated.  Hence, in the minimal substraction scheme, 
no corresponding counterterm is needed. This reflects the fact that the contact term is zero in this range.\footnote{In the Wilsonian scheme 
the contact term is an {\it irrelevant} operator in this range.} This extends to all orders in perturbation theory.

The case of the OPE~(\ref{eq:disttachope}) is different, as it leads to a logarithmic divergence for any  $r\neq 0$. 
From~\refe{eq:beta} the relevant $\beta$-functions are of the form\footnote{The sign is opposite here since the sign in front of the tachyon perturbation in~\refe{eq:act_pert} is opposite.}~: 
\eqali{
\beta_\pm  = (1-h_\pm) \lambda^\pm + (D^{\pm}_{r \pm}+ D^{\pm}_{\pm r}) \frac{\delta r}{2}  \lambda^\pm + \ldots \nonumber \\
}
We get at second order that 
\eqali{
\beta_\pm & = \left(\frac{1}{2} - r^2 - \omega^2 - 2  r \delta r\right) \lambda^\pm  \, .
}
this is valid in any scheme, as only universal quantities appear. If one keeps the distance perturbation to zero ($\delta r=0$) then the rolling tachyon background is marginal 
at second order provided that the on-shell condition $\omega^2+r^2 =1/2$, as expected. 

Otherwise, the marginality of the perturbation is restored, at this order, if we use instead the on-shell condition 
\begin{equation}\label{modonshell}
\omega^2+ (r+\delta r)^2 =1/2
\end{equation}
This is compatible with the interpretation of the boundary perturbation $\sigma^3 \otimes i \partial_u \tilde{X}$, that 
changes the relative position of the D-brane and the anti D-brane. It is T-dual to the relative Wilson line that appears 
in the action~(\ref{eq:action_begin}).\footnote{To be more correct, as auxiliary fields from the Fermi superfield couples to this perturbation, 
some $\pm i \,\delta r \, \lambda^\pm \psi^x e^{\pm i r\tilde X +\omega X^0}$ correction should be included. We verify that it doesn't modify the 
$\beta$-function at quadratic order. Moreover,
this term shows up naturally if we 
work directly with the superspace distance perturbation $i \,\delta r \, \Gam^+\Gam^-D_u{\mathbb X}$.} One checks that the  normalization of 
this coupling in~(\ref{eq:action_begin}) is compatible, through T-duality, with relation~(\ref{modonshell}). 
This analysis shows that, at least at this order, the rolling tachyon perturbations $T^{\pm}$ 'adjust themselves' to a change of 
brane-antibrane separation in order to stay marginal.

\subsubsection*{\boldmath Beta-functions for $1/2<|r|<r_c$ and contact term}
When $1/2<|r|<r_c$ the situation is different. The operator $\exp 2\omega X_0$ (that appears also in the contact term) becomes 
relevant, hence should be considered in the discussion. As stated earlier, this operator in unphysical from the superstring 
theory point of view (at zero superghost number).

The corresponding boundary coupling is denoted by $\mu_c$. The tachyon-tachyon OPE~(\ref{eq:tachtachope}) gives a singular perturbative integral at second order: 
\begin{equation}
\label{integopeMS}
\int \di x_1 
\int_{x_1-L}^{x_1-\varepsilon} \di x_2 \, \bnormal{\psi^{+}  e^{\omega X_0 + ir\tilde{X} } (x_1)}\bnormal{\psi^{-}  e^{\omega X_0 - ir\tilde{X} } (x_2)}
\stackrel{r\neq 1/2}{\sim} \varepsilon^{1-4r^2} \int \di x_1 \, \bnormal{e^{2\omega X_0}(x_1)} \, ,
\end{equation}
after removing the IR cutoff ($L \to \infty$ limit). 

In the minimal substraction scheme, the following local counterterm is needed at this order to cancel the divergence:
\begin{equation}
S_{ct} = \lambda^+ \lambda^- \varepsilon^{1-4r^2} \int \di x \, \bnormal{e^{2\omega X_0}(x)}\, .
\end{equation} 
Naturally, it agrees precisely with the expression of the contact term in the action~(\ref{eq:wscompaction}). Since this divergence is power-like, 
it does not add any non-linear term in the minimal scheme beta-function $\beta_c^\textsc{ms}$ for the coupling $\mu_c$. Hence, the latter can be 
consistently set to zero in the renormalized theory at this order. 

For the distance $|r|=1/2$, amplitudes are finite without the counterterm, so it is not strictly needed\footnote{But partition function appears to be discontinuous at 
$|r|=1/2$ without its contribution.}, but it  contributes nevertheless finitely to the amplitudes.

In the Wilsonian scheme, the beta-function  reads, at second order:
\begin{equation}
\beta_c^{\textsc{ws}} = (1-4r^2)\mu_c -  \left(1-4r^2\right)\lambda^+ \lambda^-
\end{equation}
One sees here an interesting phenomenon. The operator $\exp 2\omega X_0$ is relevant at linear order, but 
the RG flow gives an IR fixed point for this coupling at quadratic order, for $\mu_c = \lambda^+ \lambda^-$. 

Comparing the outcomes of both schemes, one gets the same results but the interpretation is different. In the minimal substraction scheme the contact term 
appears as a counterterm, but the corresponding renormalized coupling is consistently set to zero. On the contrary, in the Wilsonian scheme, the RG flow 
has a fixed point with non-zero renormalized coupling $\mu_c$. Both points of view are 'non-supersymmetric', as in the superspace formulation 
this term is present from the beginning and removes the divergence under discussion.

\subsection{Marginality beyond quadratic order}  
Part of the quadratic order results generalizes immediately to higher orders. Indeed only the fusion of  distance perturbations 
with, say, $T^+$ can produce $T^+$ itself (since the fusion of $n$ tachyons goes as $e^{n\omega X_0}$, as far as the $X_0$ dependence 
is concerned). Hence, if we set $\delta r=0$ from the very beginning, we expect that the beta-functions $\beta_\pm$ vanish to all orders 
in perturbation theory. With the same reasoning, the operator $\exp (2\omega X_0)$ that we had to consider for $|r|>1/2$ cannot receive 
higher-order contributions to its beta-function. 
 
However, study of the marginality at higher orders is quite messy when $|r|$ is getting closer to the critical distance, as the fusion of tachyon vertex 
operators produces more and more relevant  boundary operators. For a given value of $r$, these operators, of the form $e^{2n\omega X_0}$ with $n\in \mathbb{Z}_+$, 
become (superficially) relevant if $n<(2-4r^2)^{-1/2}$, and are of dimension one when they saturate this bound.  
These resonances occur all for $\nicefrac{1}{2} \leqslant |r| < \nicefrac{1}{\sqrt{2}}$; this range was excluded by  Bagchi and Sen in 
their analysis~\cite{Bagchi:2008et} for this precise reason.

A given operator $e^{2n\omega X_0}$ appears first at order $2n$ in the perturbative expansion in the tachyon perturbations, 
hence the beta-function $\beta_n$ for its coupling $\lambda_n$ is of the form: 
\begin{equation}
\beta_n = (1-4n^2 \omega^2) \lambda_n + \mathcal{O}\left( (\lambda^+ \lambda^- )^n \right)
\end{equation}
It is easier then to work in the minimal substraction scheme, where one just has to worry about logarithmic divergences, $i.e.$ 
resonances.  As we emphasized above, if the fusion of (superficially) marginal operators produces a (superficially) marginal operator, it 
generates a source term in the corresponding minimal scheme beta-function. It is nevertheless interesting to consider whether 
power-like divergences are also present.

At second order the potentially marginal operator is nothing but the contact term itself, $e^{2\omega X_0}$,  for the distance $|r|=1/2$. 
Fortunately, thanks to its fermionic part the OPE~(\ref{eq:tachtachope}) vanishes, hence there is no logarithmic divergence to cancel.

\subsubsection*{\boldmath Marginality for $\sqrt{7}/4<|r|<\sqrt{17}/6$}
The next possible resonance occurs when the operator $e^{4\omega X^0}$ becomes of dimension one, $i.e.$ for $\omega=1/4$ (equivalently, 
$|r|=\sqrt{7}/4$). The potential logarithmic divergence would occur at fourth order in perturbation theory. In order to investigate this issue we compute below 
all the possible divergent terms that occur at order $(\lambda^+ \lambda^-)^2$ from the perturbative integrals, that involve both the 
tachyon and contact term vertex operators. In the computations of this subsection, we use the full non-local contact term~\refe{eq:cont_upper}, as even the 
sub-leading terms contribute {\it a priori} to the divergences.

The first contribution comes from two contact term insertions (symbolically $CC$). Using the notations $a=4\omega^2$ and 
$T^\pm = e^{\pm i r\tilde X+\omega X_0}$, it reads
\begin{multline}
CC= \parent{\begin{array}{cc} 1 & 0 \\ 0 & 1 \end{array}} \frac{\varepsilon^{2a-2}}{4} \int \di x_1 \int_{x_1-L+\varepsilon}^{x_1-2\varepsilon} \di x_2 \,  \parent{\bnormal{ T^+(x_1+\varepsilon)T^-(x_1)} + 
\bnormal{ T^-(x_1+\varepsilon)T^+(x_1)} } \\ \times \parent{\bnormal{ T^+(x_2+\varepsilon)T^-(x_2)} + \bnormal{ T^-(x_2+\varepsilon)T^+(x_2)} }  \label{ord4a}
\end{multline}
The contact term being multiplied by the Chan-Patton identity matrix. The short-distance regularization chosen here prevents any operator to approach another one at less 
that $\varepsilon$, \emph{before} integration of the auxiliary fields. 
The most natural IR cutoff prescription is to constraint two ordered operators not to move away from each other by more that $L$, also before integration of auxiliary 
fields. One gets then
\begin{multline}
CC\sim  \parent{\begin{array}{cc} 1 & 0 \\ 0 & 1 \end{array}} \Bigg(
\frac {1}{2a + 1} \left(\frac{L}{\varepsilon}\right)^{2-2a}-\left(\frac{L}{\varepsilon}\right)^{1-2a} \\- \frac{5-6a-(2a-1) 2^{2a+2} 
\,_2 \text{F}_1(1-a,-a-\frac{1}{2};-a+\frac{1}{2};\frac{1}{4})}{4(2a+1)(2a-1)}\left(\frac{L}{\varepsilon}\right)^{1-4a} \Bigg)
\\ \ \times \ L^{4a-1} \int  \di x_1 \,\bnormal{e^{4\omega ~ X_0}}(x_1)\, .
\end{multline}

The second contribution, from two tachyons and a contact term, is more involved as one has to integrate over two operator positions, leading to various 
type of singularities. One has to be careful with path ordering of the contact term with the tachyon; we have to distinguish three contributions, 
symbolically noted CTT, TCT and TTC. One finds that the contributions of CTT and TTC are equal, 
but TCT is different. We have to sum these three contributions together. Using the notation $C(x) = \bnormal{ T^+(x+\varepsilon)T^-(x)} + \bnormal{ T^-(x+\varepsilon)T^+(x)}$, 
one has:
\begin{multline}\label{ord4b} 
CTT+TCT+TTC  =\\
 - \parent{\begin{array}{cc} 1 & 0 \\ 0 & 0 \end{array}} \frac{\varepsilon^{a-1}}{2} \Bigg( \int \di x_1 \int_{x_1-L+\varepsilon}^{x_1-\varepsilon} \di x_2 \int_{x_2-L}^{x_2-\varepsilon} \di x_3 \, 
\bnormal{C (x_1)} \bnormal{\psi^{+}  T^+ (x_2)}\bnormal{\psi^{-}  T^- (x_3)} \\ 
+ \int \di x_1 \int_{x_1-L}^{x_1-2\varepsilon} \di x_2 \int_{x_2-L+\varepsilon}^{x_2-\varepsilon} \di x_3 \, 
\bnormal{\psi^{+}T^+ (x_1)}\bnormal{C(x_2)} \bnormal{ \psi^{-}T^- (x_3)} \\ 
+ \int \di x_1 \int_{x_1-L}^{x_1-\varepsilon} \di x_2 \int_{x_2-L}^{x_2-2\varepsilon} \di x_3 \, 
\bnormal{\psi^{+}T^+ (x_1)}\bnormal{\psi^{-}T^- (x_2)} \bnormal{C (x_3)} \Bigg) 
\end{multline}
Here, the whole computation is multiplied by the upper part of the identity matrix, since $T^+$ and $T^-$ are themselves multiplied 
by $\sigma^+$ and $\sigma^-$ respectively. One should also take into account the permutated version of~\refe{ord4b} which has 
ordering $T^- T^+$ instead of $T^+ T^-$. From symmetry of the OPE's under this permutation, it contributes the same result but multiplied by the 
lower part of the identity matrix. Thus, the computation of the divergent terms gives the result, see appendix~\ref{annexe_CTT}:
\begin{multline} \label{eq:result_CTT}
 CTT+TCT+TTC \sim 
 \parent{\begin{array}{cc} 1 & 0 \\ 0 & 1 \end{array}} \Bigg[ 
-\frac{2 }{1+2a}~\parent{\frac{L}{\varepsilon}}^{2-2a} +  \frac{1}{a} \parent{\frac{L}{\varepsilon}}^{1-2a} \\ 
+ \frac{2(a-1)}{3a}   \left( \frac{L}{\varepsilon} \right)^{1-a}  
\Bigg( 
\frac{\,_2\text{F}_1\left(-a,a+1,a+2,-1\right)}{a+1} + \frac{\,_2\text{F}_1\left(-a,a-1,a,-1\right)}{a-1} \\+ \frac{\,_2\text{F}_1\left(2-a,a+1,a+2,-1\right)}{a+1}
\Bigg) 
- V(a)  \parent{\frac{L}{\varepsilon}}^{1-4a}
\Bigg]  
~ L^{4a-1}  \int \di x_1 \,\bnormal{e^{4\omega ~ X_0}}(x_1)
\end{multline}
The coefficient  $V(a)$ is given by (we did not find a closed form for it):
\begin{multline} \label{eq:Va}
V(a) =  (a-1)\sum_{n=0}^\infty \sum_{s=0}^1 \frac{\Gamma(a)}{\Gamma(a-n)\Gamma(1+n)(3a-s-n)} \Bigg( 
		\frac{\,_2 \text{F}_1 (n-a,1+n-2a;2+n-2a;-1)}{1+n-2a} \\ + \frac{\,_2 \text{F}_1 (s-a,1+s-2a;2+s-2a;-1)}{1+s-2a} 
+ \frac{\,_2 \text{F}_1 (n-a,s+n-1-2a;s+n-2a;-1)}{s+n-1-2a} \\ + \frac{\,_2 \text{F}_1 (s-a,n+s-1-2a;n+s-2a;-1)}{n+s-1-2a} \Bigg) \displaybreak[2]\\ 
+ (a-1)\sum_{n,p=0}^\infty \frac{\Gamma(a) \Gamma(a-1)}{\Gamma(a-n)\Gamma(1+n)\Gamma(a-1-p)\Gamma(1+p)} 
\frac{\,_2 \text{F}_1(1-a,n+p-3a,n+p+1-3a,-1)}{3a-n-p} \\ \times \frac{\,_2 \text{F}_1(2+p-a,n+p+1-2a,n+p+2-2a,-1)}{n+p+1-2a} \\
    + (a-1) \sum_{p=0}^\infty\sum_{s,t=0}^1 \frac{\Gamma(a-1)}{\Gamma(a-1-p)\Gamma(1+p)(3a-s-t-p)} \\ \times \ \frac{\,_2 \text{F}_1(2+p-a,s+p+1-2a,s+p+2-2a,-1)}{s+p+1-2a}
\end{multline}

Finally, one has to consider the contribution from four tachyon insertions in the path integral (TTTT). 
The method of computation of the multiple integral is explained in appendix~\ref{sec:app_TTTT}. 
After a lengthy computation one gets\footnote{The term with ordering $T^- T^+T^-T^+$ contributes the same result thus the total computation is directly multiplied by 
the identity matrix as in~\refe{eq:result_CTT}.}
\begin{multline}\label{ord4c}
TTTT = \\  \parent{\begin{array}{cc} 1 & 0 \\ 0 & 1 \end{array}} \int \di x_1 \int_{x_1-L}^{x_1-\varepsilon} \di x_2 \int_{x_2-L}^{x_2-\varepsilon} \di x_3  \int_{x_3-L}^{x_3-\varepsilon} \di x_4 \,
\bnormal{\psi^+ T^+ (x_1)}\bnormal{\psi^- T^- (x_2)}\bnormal{\psi^+ T^+ (x_3)}\bnormal{\psi^- T^- (x_4)}  \\
\sim \parent{\begin{array}{cc} 1 & 0 \\ 0 & 1 \end{array}}
\Bigg\{ 
	\frac{1}{2a+1} \parent{\frac{L}{\varepsilon}}^{2-2a} + \frac{a-1}{a} \parent{\frac{L}{\varepsilon}}^{1-2a} \\ - \frac{2(a-1)}{3a} \parent{\frac{L}{\varepsilon}}^{1-a}  
    \Bigg(
   			\frac{\,_2\text{F}_1\left(-a,a+1,a+2,-1\right)}{a+1} + \frac{\,_2\text{F}_1\left(-a,a-1,a,-1\right)}{a-1}\\
\hphantom{aaaaaaaaaaaaaaaaaAAaaaaaaaaaaaaaaaaaaaaaaa} + \frac{\,_2\text{F}_1\left(2-a,a+1,a+2,-1\right)}{a+1} 
    \Bigg) \displaybreak[3]\\ 
   + \frac{\parent{\frac{L}{\varepsilon}}^{1-4a}-1}{1-4a} 
	\Bigg[ 
		 	(a-1)^2  \parent{\frac{\,_2\text{F}_1\left(1-2a,a-1,a,-1\right)}{a-1} + \frac{\,_2\text{F}_1\left(1-2a,2-3a,3-3a,-1\right)}{2-3a}} \\ \times 
		 	\Bigg(  
		 			\frac{\,_2\text{F}_1\left(-a,a-1,a,-1\right)}{a-1} + \frac{\,_2\text{F}_1\left(-a,1-2a,2-2a,-1\right)}{1-2a} \\ 
+ \frac{\,_2\text{F}_1\left(2-a,a+1,a+2,-1\right)}{a+1}+ \frac{\,_2\text{F}_1\left(2-a,1-2a,2-2a,-1\right)}{1-2a} 
		 	\Bigg) \\ 
			+ \parent{2(a-1)^2-1}  \parent{\frac{\,_2\text{F}_1\left(1-a,a,1+a,-1\right)}{a}
+\frac{\,_2\text{F}_1\left(1-a,1-2a,2-2a,-1\right) }  {1-2a}}    \\ \times \Bigg( \frac{\,_2\text{F}_1\left(1-2a,a,a+1,-1\right)}{a} 
+ \frac{\,_2\text{F}_1\left(1-2a,1-3a,2-3a,-1\right)}{1-3a} \Bigg)
	\Bigg] \\ 
+ U(a) \parent{\frac{L}{\varepsilon}}^{1-4a} 
\Bigg\}  
~ L^{4a-1}  \int \di x_1 \,\bnormal{e^{4\omega ~ X_0}}(x_1)
\end{multline}
with $U(a)$ a numerical coefficient which is not singular at $a=1/4$.

As in the previous computation, the coefficient $U(a)$ is known only as a series expansion 
\begin{multline} \label{eq:Ua}
U(a) = \frac{(a-1)^2 }{4 a-1}\left(\frac{\, _2\text{F}_1(1-2 a,a-1;a;-1)}{a-1}+\frac{\, _2\text{F}_1(1-2 a,-3 a;1-3 a;-1)}{3 a}\right) \\ 
\times \Bigg(\frac{\, _2\text{F}_1(-a,1-2 a;2-2 a;-1)}{1-2 a}+\frac{\, _2\text{F}_1(-a,a-1;a;-1)}{a-1} \\ + 
\frac{\, _2\text{F}_1(2-a,1-2 a;2-2 a;-1)}{1-2 a}+\frac{\, _2\text{F}_1(2-a,a+1;a+2;-1)}{a+1} \Bigg)  \\
\displaybreak[3]+ \frac{\left(2 (a-1)^2-1\right)}{4 a-1} \left(\frac{\, _2\text{F}_1(1-2 a,1-3 a;2-3 a;-1)}{3 a-1}+\frac{\, _2\text{F}_1(1-2 a,a;a+1;-1)}{a}\right)\\ 
\times \left(\frac{\, _2\text{F}_1(1-a,1-2 a;2-2 a;-1)}{1-2 a}+\frac{\, _2\text{F}_1(1-a,a;a+1;-1)}{a}\right) \\
\displaybreak[2]+ (a-1)^2\sum _{n=0}^{\infty } \frac{ \Gamma (a+1) }{ \Gamma (n+1) \Gamma (a-n+1)(a+n-1)(3 a-n) }\\ 
\left(\frac{\, _2\text{F}_1(n-a,-2 a+n+1;-2 a+n+2;-1)}{-2 a+n+1}+\frac{\, _2\text{F}_1(n-a,-2 a+n-1;n-2 a;-1)}{-2 a+n-1}\right) \\ 
+ (a-1)^2\sum _{n=0}^{\infty } \frac{ \Gamma (a-1) }{ \Gamma (n+1) \Gamma (a-n-1)(a+n+1)(3 a-n-2) } \\ 
\Bigg(\frac{\, _2\text{F}_1(-a+n+2,-2 a+n+1;-2 a+n+2;-1)}{-2 a+n+1}\\
\hphantom{aaaaaaaa}+\frac{\, _2\text{F}_1(-a+n+2,-2 a+n+3;-2 a+n+4;-1)}{-2 a+n+3}\Bigg)\displaybreak[3] \\
+ 2 \left(2 (a-1)^2-1\right)  \sum _{n=0}^{\infty } \frac{\Gamma (a) }{\Gamma (n+1) \Gamma (a-n)(a+n)(3 a-n-1)} \\ \frac{\, _2\text{F}_1(-a+n+1,-2 a+n+1;-2 a+n+2;-1)}{-2 a+n+1}
\end{multline}
The last three sums are rapidly converging, thus $U(a)$ is known with good accuracy, for any value of $a \leqslant 1/4$ (or $\omega\leqslant 1/4$).

Let us now investigate the possible logarithmic divergences, that can only occur from the $TTTT$ integral. Since we have that  
\begin{equation}
\frac{\parent{\frac{L}{\varepsilon}}^{1-4a}-1}{1-4a} \stackrel{a\to 1/4}{\to} \log \frac{L}{\varepsilon}\, ,
\end{equation}
only the last but one term in~(\ref{ord4c}) could lead to a logarithmic divergence at $\omega=1/4$. It turns out that, in this limit,  
the coefficient of this term vanishes exactly.  Looking more closely at this computation, one sees 
that each multiple integral that one gets from the three different fermionic contractions --~see eq.~(\ref{eq:4thordrer})~--
has a logarithmic term as expected, however the sum of them precisely cancels. Hence, the same occurs as at order two; the coefficient in front of the 
potentially resonant term in the beta-function vanishes.\footnote{This is confirmed by a direct evaluation of the TTTT integral at $\omega=1/4$ 
(with \textsc{Mathematica}) which gives 
\begin{multline*}
TTTT =  \parent{\begin{array}{cc} 1 & 0 \\ 0 & 1 \end{array}}  4 \int \di x_1 \int_{x_1-L}^{x_1-\varepsilon} \di x_2 \int_{x_2-L}^{x_2-\varepsilon} \di x_3  \int_{x_3-L}^{x_3-\varepsilon} \di x_4 \,
\bnormal{\psi^{+}  e^{ X_0/4 + ir\tilde{X} } (x_1)}\bnormal{\psi^{-}  e^{X_0/4 - ir\tilde{X} } (x_2)}\ \times \\ \ \times \ 
\bnormal{\psi^{+}  e^{X_0/4 + ir\tilde{X} } (x_3)}\bnormal{\psi^{-}  e^{X_0/4 - ir\tilde{X} } (x_4)} \\
\sim \left[ \frac{2}{3}\left(\frac{L}{\varepsilon}\right)^{3/2}+
\left(\frac{7  \sqrt{\pi } \Gamma\left(\frac{5}{4}\right)}{3  \Gamma\left(\frac{3}{4}\right)}-\alpha\right)\left(\frac{L}{\varepsilon}\right)^{3/4}
-3 \left(\frac{L}{\varepsilon}\right)^{1/2}
\right] \int\,  \di x_1 \, \bnormal{e^{X_0} (x_1)}  
\end{multline*}
with $\alpha\simeq 1.24\ldots$. Logarithmic divergences are again found to vanish.} 

In order to check whether power-like divergences remain at fourth order, one has to resum the three contributions obtained above. The full contribution 
at  order $\parent{(\lambda^+\lambda^-)^2}$ is given by  $CC + CTT + TCT + TTC + TTTT$. Comparing~\refe{ord4a},~\refe{ord4b} and~\refe{ord4c}, one sees 
that the coefficients in front of all divergent terms vanish exactly for any value of $\omega \geqslant 1/4$. Hence, in this range, if one includes the two-tachyon 
contact term dictated by worldsheet supersymmetry, perturbative expansion is  finite.\footnote{
Not considering into account possible operator renormalization if there are operator insertions in the path integral.}

As said before we were not able to compute  the coefficient associated to the term of order $\epsilon^{1-4a}$, which becomes divergent 
for $\omega<1/4$ in a closed form. Using  a numerical evaluation, we find that the sum of the contributions gives 
a non-zero coefficient for any  $\omega < 1/4$. Hence, a power-like divergence remains in this range.
By dimensional counting, this uncanceled divergence corresponds to four tachyon operators coming close together at the same point. It is not 
unexpected that this divergence is not canceled by the contact term, as the latter corresponds to a two-tachyon collision. Since this 
remaining divergence is non-logarithmic, it does not mean that the boundary theory is not conformal, but rather that it should be renormalized 
at quartic order. It should be possible to cancel this divergence with  higher-order contact-term. They may correspond to additional non-linear terms 
in the superspace action~(\ref{eq:action}) (a four-auxiliary field vertex is needed then).

As mentioned in section~\ref{sec:dist_pert}, we also obtained an unexpected result. If we assume that the computations of CTT and CC type terms could be 
equivalently done with the use of the simple dominant term $\varepsilon^{a-1} e^{2\omega X^0}$ in~\refe{eq:cont_upper}, then we get the following contribution 
\begin{multline} \label{eq:counter}
CC + CTT+TCT+TTC =  \parent{\begin{array}{cc} 1 & 0 \\ 0 & 1 \end{array}} \Bigg\{ -\frac{1 }{1+2a}~\parent{\frac{L}{\varepsilon}}^{2-2a} +  
\frac{1}{a} \parent{\frac{L}{\varepsilon}}^{1-2a} \\ 
+ \frac{2(a-1)}{3a}   \left( \frac{L}{\varepsilon} \right)^{1-a}  
\Bigg( 
\frac{\,_2\text{F}_1\left(-a,a+1,a+2,-1\right)}{a+1} + \frac{\,_2\text{F}_1\left(-a,a-1,a,-1\right)}{a-1} \\ + \frac{\,_2\text{F}_1\left(2-a,a+1,a+2,-1\right)}{a+1}
\Bigg) \\  - \left( \frac{L}{\varepsilon} \right)^{1-4a} \Bigg[ 2\frac{a-1}{3a} \Bigg( \frac{\, _2F_1(2-a,1-2a;2-2a;-1)}{1-2a} +  \frac{\, _2F_1(-a,1-2a;2-2a;-1)}{1-2a} \\ - 
\frac{\, _2F_1(-a,-1-2a;-2a;-1)}{1+2a} \Bigg) + \frac{1}{2a+1} \Bigg] \Bigg\} ~ L^{4a-1}  \int \di x_1 \,\bnormal{e^{4\omega ~ X_0}}(x_1)
\end{multline} 
One recognizes the coefficients of the three first divergences; these are precisely the ones appearing in the sum of~\refe{ord4a} and~\refe{ord4b}. Moreover, numerical
comparison of the $\left( \frac{L}{\varepsilon} \right)^{1-4a} $ coefficients gives almost identical results; the tiny difference could reasonably originates from the 
approximated evaluation of the infinite sums. This seems to show the equivalence of the two computations --~\refe{eq:counter} being of course significantly easier to 
perform~-- and then of the two (local and non-local) expressions of the contact term.

\subsubsection*{Marginality to all orders}

Computations become intractable for the next resonance, which occurs for $\omega =1/6$ (or equivalently $|r|=\sqrt{17}/6$), as we have to consider  sixth order perturbation theory, 
with contributions from both counterterms found so far. However we assume that the same occurs; the coefficient in front of 
the logarithmic six-tachyon divergence should vanish as well. 

To summarize, we have found that, to all orders in perturbation theory, the theory defined by the boundary action~(\ref{eq:wscompaction}) is a boundary 
conformal field theory when $|r|<\sqrt{17}/6$. In the range $\sqrt{17}/6<|r|<1/\sqrt{2}$, the theory is  conformal at least up to order five. 
We naturally expect that the theory is conformal to all orders in this range as well.

As a side remark, the theory defined by the limit $r\to r_c^{\, -}$ seems not well-defined. In this case, all the operators $e^{2n\omega X_0}$ are relevant, 
and by doing the perturbative expansion in the tachyon couplings we would need an infinite number of counter-terms. By contrast, the theory defined 
directly at $r=r_c=1/\sqrt{2}$ seems fine. The boundary interaction (with $\mathbb{T}^\pm\sim e^{\pm i \mathbb{X}/\sqrt{2}}$) 
is similar to a boundary sine-Gordon theory, with additional CP factors. Other puzzling features of the $r\to r_c^{\, -}$ limit will be discussed 
in the next section.

\section{Discussion}
\label{sec:effective}
We argued in this work that, for all values of the brane-anti-brane distance below the critical value $r_c$, the homogeneous rolling 
tachyon solution with a fixed separation is an exact boundary conformal field theory. Thus, a spacetime effective action that is 
valid around this particular solution should have such tachyon profile as a solution 
of its equations of motion. An effective action for the brane-anti-brane system was proposed by Garousi~\cite{Garousi:2004rd}. In a different 
parameterization of the tachyon field,\footnote{This field redefinition was discussed in~\cite{Kutasov:2003er} for the $r=0$ case.}, 
it reads:
\begin{equation}\label{Garousi}
\mathcal{L}_{\textsc{g}} (T,\dot{T},r,0) = - \frac{2}{\cosh \sqrt{\pi |T|}} \sqrt{1+4\pi^2 r^2 |T|^2 -|\dot{T}|^2-\pi^2 \dot{r}^2}
\end{equation}
One checks readily that, with $\dot r=0$, $\delta_r \mathcal{L}_{\textsc{g}} \neq 0$ for any non-zero separation. Hence, this Lagrangian cannot admit solutions with constant 
brane-antibrane separation. This is not to be unexpected, since it was obtained by a fermion number orbifold of the non-Abelian tachyon-DBI action for 
a pair of coincident non-BPS D-branes. Therefore it could only be valid for an infinitesimal brane separation. 
Since  $\delta_r \mathcal{L}_{\textsc{g}}$ is linear in $r$, it seems not even to be valid in this limit. 

In order to find the space-time effective action from first principles, we could proceed as in~\cite{Kutasov:2003er}. In this approach, one considers 
a generic spacetime Lagrangian of the D0-$\bar{\text{D}}$0 system, depending on the tachyon field $\tau$, its first derivative, the distance field $r$ 
and its first derivative.\footnote{We assume that, 
by the symmetries of the problem, only even powers of the fields and their derivative appear, $i.e$  one has $\mathcal{L} (|\tau|^2, |\dot \tau|^2, r^2, \dot r^2)$. 
Without loss of generality, as the phase of the tachyon for the 'half S-brane' solution under study is constant, we take $\tau(t)$ real.}  Since, 
as we argued before, rolling tachyon solutions at constant separation exist, the effective  Lagrangian  describing nearby field configurations should satisfy
the condition 
\begin{equation}
\label{eq:eomst}
\frac{\delta \mathcal{L} (\tau,\dot \tau,r,\dot r)}{\delta r} \Big|_{\dot{r}=0, \dot{\tau}= \omega \tau} = 0
\end{equation}
where  $\omega^2=\tfrac{1}{2}-r^2$, as well as the equation of motion for the tachyon with a  profile of the form  $\tau=\mu \exp {\omega t}$.  

Solving these equations at quadratic order in the tachyon field, one obtains a unique result,  if we ask that for $r=0$ one should recover 
the known Lagrangian for the coincident case: 
\eqali{\label{secondorder}
{\mathcal L} (\tau,\dot \tau,r,0)= -2 + \sqrt{1-2r^2} \left( \frac{\tau^2}{2} + \frac{\dot{\tau}^2}{1-2r^2}\right) + \cdots
}

Unlike in the case of the non-BPS brane considered by Kutasov and Niarchos in~\cite{Kutasov:2003er}, we did not impose above that the more generic profile  
$\tau=\zeta e^{\omega t}+ \xi e^{-\omega t}$ (with arbitrary coefficients $\zeta$ and $\xi$) is a solution, since  it does not correspond to an exactly marginal 
deformation on the worldsheet as long as $r\neq 0$. A straightforward generalisation of the effective Lagrangian found by these authors exists for 
$r\neq 0$but should not be considered since, by construction, it allows  the time-reversal-symmetric tachyon profile $\tau \sim \cosh \omega t$ as a solution.

Imposing only the 'half S-brane' as a solution leads to an underconstrained (finite) system of equations and not to a single recurrence relation as in~\cite{Kutasov:2003er}.  
However, below eq.~\refe{generalbetadistance}  we have shown that a solution of this form is marginal at second order, provided $\xi = i \mu \zeta$ with $\mu$ real. 
Besides the necessity to prove its conformal invariance at all order (by going through an even more tedious analysis as we have done for the 'half S-brane' solution), it would again 
lead to an underconstrained system. Indeed, this tachyon satisfies the identity $\module{\dot\tau}^2 = \omega^2 \module{\tau}^2$. One can show that, as a consequence, the relation 
between the coefficients in the Lagrangian does not organize into a single recurrence relation, but rather separates into a  system of independent equations 
which is underconstrained.   Thus it does not lead to a unique  effective Lagrangian at higher orders in the tachyon couplings.\footnote{Note on the other hand that~\refe{secondorder} is 
still valid with~\refe{fullSbrane} under the replacement $\tau^2 \to \module{\tau}^2$ and $\dot\tau^2 \to |\dot \tau|^2$.}

The worldsheet theory contains more information about the tachyon effective action, besides imposing that the rolling tachyon background 
of interest should be a solution of its equations of motion. Following~\cite{Kutasov:2003er,Niarchos:2004rw}, we expect to get the effective Lagrangian 
evaluated {\it on-shell} to be given by the disk partition function, with the time-like zero modes kept unintegrated:
\begin{equation}
\label{eq:onshellrel}
\mathcal{L}\big|_{\tau,\omega \tau,r,0} (x_0) = -Z (r|x_0)\big|_{\text{disk}}
\end{equation}
With this equation one can  test whether any proposal for the effective Lagrangian of the system is sensible. At second order, one can compare the spacetime Lagrangian, 
given by eq.~(\ref{secondorder}), with the partition function given in appendix~\ref{app:partition}:
\eqali{
Z(r|x_0) = 2 - \frac{\Gamma(2-4r^2)}{\Gamma^2(1-2r^2)} \lambda^+\lambda^- e^{2\omega x_0} + 
\mathcal{O}\Big((\lambda^+ \lambda^-)^2\Big)   \label{eq:part_order_1_bulk}
}
In order to match these two computations, we see that a distance-dependent 
field redefinition of the tachyon field is necessary:
\eqali{\label{eq:fieldredef}
\tau (t) = \left(\frac{1}{\sqrt{1-2r^2} }\frac{\Gamma(2-4r^2)}{\Gamma^2(1-2r^2)}\right)^{1/2}\, \lambda^+ e^{\omega t}
}
As one can see, with this definition, the spacetime tachyon vanishes at the critical distance $r=1/\sqrt 2$, for any finite value of the 
worldsheet coupling $\lambda^+$. It could be the way string theory deals with the fact that, when $r\to r_c$, the tachyon becomes a light field, and 
we could wonder how a {\it local} action along the brane worldvolume dimensions --~that is {\it a priori}  well-defined  as the tachyon is lighter than all 
string modes~-- would make sense, since the separation between the brane and the antibrane is significant in this regime.

The validity of the field redefinition~(\ref{eq:fieldredef}) should be tested beyond quadratic order.\footnote{One can check 
already that plugging this redefinition in Garousi's Lagrangian~(\ref{Garousi}) does not lead to a consistent effective Lagrangian.} For this one would have  first 
to compute analytically the perturbative 'screening integrals' at higher order which does not seem trivial. 
For the special value of the distance $r=1/2$ the computation, up to order eight in the tachyon amplitude, is given  in appendix~\ref{app:partition}:
\begin{multline}\label{eq:8order}
Z (\tfrac{1}{2}|x_0) = 2\Bigg( 1- \frac{\lambda^+ \lambda^- e^{\omega x_0}}{2\pi}  + 
\left(1-\frac{\pi^2}{6}\right) \left(\frac{\lambda^+ \lambda^- e^{\omega x_0}}{2\pi}\right)^2
- \left(1-\frac{128}{3\pi^2}\right) \left(\frac{\lambda^+ \lambda^- e^{\omega x_0}}{2\pi}\right)^3 \\
+ \left(1+\frac{205}{108}+\frac{3575}{162 \pi ^2}+\frac{\pi ^2}{2}+\frac{\pi ^4}{70} \right)\left(\frac{\lambda^+ \lambda^- e^{\omega x_0}}{2\pi}\right)^4 
+ \mathcal{O}\left( \left(\frac{\lambda^+ \lambda^- e^{\omega x_0}}{2\pi}\right)^5 \right) \Bigg)
\end{multline}
This does not seem to trace back to the Taylor expansion of a known function.  

Since the space-time effective action approach seems to have important limitations for branes-antibranes at finite separation, the  boundary string field theory may 
be more appropriate in order to know the properties of the system. Even though it does not contain 
information about the dynamics of the system, it allows to find the exact tachyon potential (as well as the appearance of lower-dimensional branes), 
hence can illuminate the fate of the tachyon.  These computations seem not to be out of reach. We plan to come back to these issues in the near future.

A heuristic argument gives a good motivation for this study.  Following ~\cite{Lambert:2003zr,Sen:2002in}, one could describe the 
result of this condensation by studying the closed string emission from the time-dependent boundary state. 
It was found in~\cite{Lambert:2003zr} that, knowing the one-point function on the disk $B(E)=\langle e^{iE X_0} \rangle$, one can compute 
the density of closed string states emitted by the decay of a non-BPS brane which goes as 
\begin{equation}
\rho_{c} \sim \sum_N \frac{1}{E_N} D(N) |B(E_N)|^2  
\end{equation}   
where the asymptotic Hagedorn density of closed string states at level $N$ has the form $D(N)\sim N^{-\alpha} \exp (4\pi\sqrt{N})$, with $\alpha >0$, and 
$E_N \sim 2\sqrt{N}$. The one-point function for an unstable non-BPS D-brane goes as $|B(E)|^2 \sim \exp (-2\pi E )$. 
Therefore, in this case, the sum is governed by the sub-leading power-like corrections to the Hagedorn density and typically diverge, giving the so-called 'tachyon dust' of 
massive closed strings. In the case of non-zero separation, by dimensional analysis we may expect that  $|B(E)| \sim \exp (-\sqrt{2}\pi E/ |m_{tach} (r)|)$. This  
would lead to a convergent closed string production when $|m_{tach}|<1/\sqrt{2}$ ($i.e.$ for $r\neq 0$), 
signaling that the tachyon does not condense completely at finite distance.

\section{Acknowledgements}
We thank Vladimir Dotsenko, Matthias Gaberdiel, Costas Kounnas,  Vasilis Niarchos and Jan Troost for stimulating discussions. This work is in part supported by the ANR grant 
STR-COSMO, ANR-09-BLAN-0157.

\appendix

\section{Computation of the divergences in CTT-type terms} \label{annexe_CTT}
We give below one example of computation of the divergence occuring in an integral involving one contact operator insertion. 
We study here the CTT term. With a bit of care, one can compute them exactly. With the expression of $C$ given in~\refe{eq:cont_upper}, the CTT term is 
\begin{multline}
- \frac{\varepsilon^{a-1}}{2} \int \di x_1 \int_{x_1-L+\varepsilon}^{x_1-\varepsilon} \di x_2 \int_{x_2-L}^{x_2-\varepsilon} \di x_3 \, 
\bnormal{C (x_1)} \bnormal{T^+ (x_2)}\bnormal{T^- (x_3)} \\ 
= (a-1) \frac{\varepsilon^{a-1}}{2} \int \di x_1 \int_{x_1-L+\varepsilon}^{x_1-\varepsilon} \di x_2 \int_{x_2-L}^{x_2-\varepsilon} \di x_3 \,  
\Bigg( (x_1-x_2 + \varepsilon) (x_1-x_3 + \varepsilon)^{a-1} (x_1-x_2)^{a-1} (x_1-x_3) (x_2-x_3)^{a-2} \\ + (x_1-x_2 + \varepsilon)^{a-1} (x_1-x_3 
+ \varepsilon) (x_1-x_2) (x_1-x_3)^{a-1} (x_2-x_3)^{a-2} \Bigg) \\
\end{multline}
with $a=4\omega^2$. Note that the IR cut-off is chosen such that two ordered operator do not move away from each other more that $L$. 
Then, since $C(x) \sim T^\pm (x+\varepsilon) T^\mp(x)$ the cut-off for $x_2$ in relation to $x_1$ is $L-\varepsilon$. One can get read of the path ordering with 
the following change of variable : 
\eqali{
x_2 = - L \delta_1 + x_1 \nonumber\\
x_3 = - L \delta_2 + x_2   
}
such that it gives, introducing $\eta = \varepsilon/L$: 
\begin{multline} 
(a-1) L^{4a-1} \frac{\varepsilon^{a-1}}{2} \int^{1-\eta}_{\eta}  \di \delta_1 \int^{1}_{\eta} \di \delta_2  \, 
\Bigg(  (\delta_1+\eta) (\delta_1+\delta_2+\eta)^{a-1} \delta_1^{a-1} (\delta_1+\delta_2) \delta_2^{a-2}  \\  
 + (\delta_1+\eta)^{a-1} (\delta_1+\delta_2+\eta) \delta_1 (\delta_1+\delta_2)^{a-1} \delta_2^{a-2} \Bigg)\int \di x_1 ~ e^{4\omega X^0}(x_1)\\
\end{multline}
The integral over $\delta_i$'s can be done with the use of the series representation of
$(1+\frac{\eta}{\delta_1+\delta_2})^\alpha$ since $\delta_1+\delta_2 >\eta$, and $(1+\frac{\eta}{\delta_1})^\beta$ since $\delta_1 >\eta$. These are given by: 
\begin{equation}
(1+x)^\alpha = \sum_{n=0}^\infty \frac{\Gamma(1+\alpha) ~ x^n}{\Gamma(1+\alpha-n)\Gamma(1+n)} \qquad \text{with   } \module x<1
\end{equation}

Convergence of the series all along the domain of integration allows us to commute integral and sum sign\footnote{It is true at least \emph{a fortiori} from the 
convergence of the integrals and the series of the integrals. Note besides that we do not integrate over any pole.}, such that one has: 
\begin{multline} \label{eq:begin_CTT}   
(a-1) \sum_{s = 0}^{1} \sum_{n=0}^\infty \frac{\Gamma(a)}{\Gamma(a-n)\Gamma(1+n)} ~ \eta^{a-1+s+n} \\ \times \int^{1-\eta}_{\eta}  
\di \delta_1 \int^{1}_{\eta} \di \delta_2  \, \Bigg( \delta_1^{a-s} \delta_2^{a-2} (\delta_1+\delta_2)^{a-n}  + \delta_1^{a-n} \delta_2^{a-2} (\delta_1+\delta_2)^{a-s} \Bigg)
\end{multline}
As one can see, the two integral to compute are symmetric by permutation of $s$ and $n$. We then only focus on the first one. There are 
two ways to proceed now. Integrate directly and exactly since it is possible, or use an indirect method that reintroduce some path ordering. 
We use the second and apparently more complicated method, because it is needed to compute TTTT integrals. Indeed, one will see that hypergeometric 
functions will receive argument $z$ which has absolute value less than $1$, much more easier to handle for approximations, since the series 
representation is known exactly. We separate the first integral of~\refe{eq:begin_CTT} into:
\begin{multline}
\int^{1-\eta}_{\eta} \di \delta_1 \int^{\delta_1}_{\eta} \di \delta_2 ~ \delta_1^{2a-n-s} \delta_2^{a-2} (1+ \frac{\delta_2}{\delta_1})^{a-n}  
+  \int^{1-\eta}_{\eta} \di \delta_1 \int^{1}_{\delta_1} \di \delta_2 ~ \delta_1^{a-s} \delta_2^{2a-2-n} (1+ \frac{\delta_1}{\delta_2})^{a-n} \\
= \int^{1-\eta}_{\eta} \di \delta_1 ~ \delta_1^{2a-n-s} \Bigg[ \frac{\delta_2^{a-1}}{a-1} 
\,_2\text{F}_1\left(n-a,a-1,a,-\frac{\delta_2}{\delta_1}\right) \Bigg]_\eta^{\delta_1 } \\
 + \int^{1-\eta}_{\eta} \di \delta_1 ~ \delta_1^{a-s} \Bigg[ \frac{\delta_2^{2a-1-n}}{2a-1-n} 
\,_2\text{F}_1\left(n-a,1+n-2a,2+n-2a,-\frac{\delta_1}{\delta_2}\right) \Bigg]_{\delta_1}^1 \\
=\int^{1-\eta}_{\eta} \di \delta_1 ~ \delta_1^{3a-1-s-n} 
\parent{ \frac{\,_2\text{F}_1\left(n-a,a-1,a,-1\right)}{a-1} + \frac{\,_2\text{F}_1\left(n-a,1+n-2a,2+n-2a,-1\right)}{1+n-2a}} \\
 - \frac{\eta^{a-1}}{a-1} \int^{1-\eta}_{\eta} \di \delta_1 ~ \delta_1^{2a-s-n}  \,_2\text{F}_1\left(n-a,a-1,a,-\frac{\eta}{\delta_1}\right) \\
 + \frac{1}{2a-1-n} \int^{1-\eta}_{\eta} \di \delta_1 ~ \delta_1^{a-s} \,_2\text{F}_1\left(n-a,1+n-2a,2+n-2a,-\delta_1 \right) \\
\end{multline}

Let us remark at this stage that $z$ argument in $\,_2 \text{F}_1(a,b,c,z)$ verifies $\module{z}<1$ in the above integrals. The first one is trivial and gives: 
\begin{multline}
I_1 = \frac{(1-\eta)^{3a-s-n}-\eta^{3a-s-n}}{3a-s-n} \parent{ \frac{\,_2\text{F}_1\left(n-a,a-1,a,-1\right)}{a-1} + \frac{\,_2\text{F}_1\left(n-a,1+n-2a,2+n-2a,-1\right)}{1+n-2a}} \\
=  \frac{1-\eta^{3a-s-n}}{3a-s-n} \parent{ \frac{\,_2\text{F}_1\left(n-a,a-1,a,-1\right)}{a-1} + \frac{\,_2\text{F}_1\left(n-a,1+n-2a,2+n-2a,-1\right)}{1+n-2a}} + ~ o(\eta) 
\end{multline}
The second one is a bit more involved
\begin{multline}
I_2 = -\frac{\eta^{3a-s-n}}{a-1} \Bigg[ -\frac{\delta_1^{s+n-1-2a}(a-1)}{3a-s-n} \Bigg( \frac{\,_2\text{F}_1\left(n-a,a-1,a,-\delta_1\right)}{a-1} ~ - \\ \frac{\,_2\text{F}_1\left(n-a,s+n-1-2a,s+n-2a,-\delta_1\right)}{s+n-1-2a}\Bigg)\Bigg]_\frac{\eta}{1-\eta}^{1}  \\
= \frac{\eta^{3a-s-n}}{3a-s-n}\Bigg( \frac{\,_2\text{F}_1\left(n-a,a-1,a,- 1\right)}{a-1} ~ - \frac{\,_2\text{F}_1\left(n-a,s+n-1-2a,s+n-2a,-1\right)}{s+n-1-2a} \Bigg)  \\
- \frac{\eta^{a-1}(1-\eta)^{-s-n+1+2a}}{3a-s-n} \parent{ \frac{\,_2\text{F}_1\left(n-a,a-1,a,-\frac{\eta}{1-\eta}\right)}{a-1} - \frac{\,_2\text{F}_1\left(n-a,s+n-1-2a,s+n-2a,-\frac{\eta}{1-\eta}\right)}{s+n-1-2a}} \\
= \frac{\eta^{3a-s-n}}{3a-s-n}\Bigg( \frac{\,_2\text{F}_1\left(n-a,a-1,a,- 1\right)}{a-1} ~ - \frac{\,_2\text{F}_1\left(n-a,s+n-1-2a,s+n-2a,-1\right)}{s+n-1-2a} \Bigg) \\
+ \frac{\eta^{a-1}}{(a-1)(s+n-1-2a)} + \frac{\eta^a }{a-1} \Bigg( 1 + \frac{(a-n)(a-1)}{a(s+n-2a)} \Bigg) + o(\eta^{a+1})
\end{multline}
On the last line we used the series representation of $\,_2\text{F}_1$  : 
\begin{equation}
\,_2\text{F}_1(a, b, c, z) = \sum_{k=0}^\infty \frac{(a)_k (b)_k  z^k}{(c)_k k!}
\end{equation}
for $\module{z}<1$. In particular, for $c=b+1$ we have: 
\begin{equation} \label{eq:series_F}
\,_2\text{F}_1(-a, b, b+1,-z) = \sum_{k=0}^\infty \frac{\Gamma(1+a) \, b }{\Gamma(1+a-k)\Gamma(1+k)(b+k)} ~ z^k
\end{equation}
Finally, the third one is : 
\begin{multline}
I_3 = \frac{1}{2a-1-n} \Bigg[- \frac{\delta_1^{a+1-s}(1+n-2a)}{s+n-3a} \Bigg( \frac{\,_2\text{F}_1\left(n-a,1+n-2a,2+n-2a,-\delta_1\right)}{1+n-2a} ~ -  \\ \frac{\,_2\text{F}_1\left(n-a,1+a-s,2+a-s,-\delta_1\right)}{1+a-s}\Bigg)\Bigg]_\eta^{1-\eta} \\
=  \frac{(1-\delta_1)^{a+1-s}}{s+n-3a} \Bigg( \frac{\,_2\text{F}_1\left(n-a,1+n-2a,2+n-2a,-1+\eta \right)}{1+n-2a} ~ - \frac{\,_2\text{F}_1\left(n-a,1+a-s,2+a-s,- 1+\eta \right)}{1+a-s}\Bigg) \\
- o(\eta^{a+1-s}) \\
= \frac{-1}{3a-s-n} \Bigg( \frac{\,_2\text{F}_1\left(n-a,1+n-2a,2+n-2a,-1 \right)}{1+n-2a} ~ -  \frac{\,_2\text{F}_1\left(n-a,1+a-s,2+a-s,- 1 \right)}{1+a-s}\Bigg) \\
+ o(\eta^{a+1-s}) + o(\eta) 
\end{multline}

Collecting these results one finally get the sum: 
\begin{multline}
\frac{a-1}{2} \sum_{s = 0}^{1} \sum_{n=0}^\infty \frac{\Gamma(a)}{\Gamma(a-n)\Gamma(1+n)} ~ \eta^{a-1+s+n} \int^{1-\eta}_{\eta}  \di \delta_1 \int^{1}_{\eta} \di \delta_2  \, \Bigg( \delta_1^{a-s} \delta_2^{a-2} (\delta_1+\delta_2)^{a-n}  + \delta_1^{a-n} \delta_2^{a-2} (\delta_1+\delta_2)^{a-s} \Bigg) \\ 
= \frac{a-1}{2} \sum_{s = 0}^{1} \sum_{n=0}^\infty \frac{\Gamma(a)}{\Gamma(a-n)\Gamma(1+n)} ~ \eta^{a-1+s+n} \parent{I_1+I_2+I_3 + (s \leftrightarrow n)}  \\
\sim - \frac{\eta^{2a-2}}{2a+1} + \frac{\eta^{2a-1}}{2a}  \\ 
- \eta^{a-1}  \frac{a-1}{3a}\parent{\frac{\,_2\text{F}_1\left(-a,a+1,a+2,-1\right)}{a+1} + \frac{\,_2\text{F}_1\left(-a,a-1,a,-1\right)}{a-1}} \\
- \sum_{n=0}^\infty \sum_{s=0}^1 \frac{\Gamma(a)}{\Gamma(a-n)\Gamma(1+n)(3a-s-n)} \Bigg(\frac{\,_2\text{F}_1\left(n-a,1+n-2a,2+n-2a,- 1 \right)}{1+n-2a} \\ +\frac{\,_2\text{F}_1\left(s-a,1+s-2a,2+s-2a,- 1 \right)}{1+s-2a}  + \frac{\,_2\text{F}_1\left(n-a,s+n-1-2a,s+n-2a,- 1 \right)}{s+n-1-2a} \\ + \frac{\,_2\text{F}_1\left(s-a,s+n-1-2a,s+n-2a,- 1 \right)}{s+n-1-2a}\Bigg)
\end{multline}

A similar computation was done for the $TCT$ and $TTC$ terms, with the correct cut-off prescriptions. Note however that $CTT=TTC$.  

\section{Computation of the divergences in the TTTT  term} \label{sec:app_TTTT}

The computation of an amplitude with four tachyon insertions is clearly a lot more involved than the above one, since three integrations have to be done. 
The straightforward OPE of the four tachyons is doable and gives, from~\refe{eq:4thordrer} : 
\begin{multline}
\int \di x_1 \int_{x_1-L}^{x_1-\varepsilon} \di x_2 \int_{x_2-L}^{x_2-\varepsilon} \di x_3  \int_{x_3-L}^{x_3-\varepsilon} \di x_4 \,
\bnormal{\psi^+ T^+ (x_1)}\bnormal{\psi^- T^- (x_2)}\bnormal{\psi^+ T^+ (x_3)}\bnormal{\psi^- T^- (x_4)} \\ 
 = \int \di x_1 \, e^{4\omega X^0} \int_{x_1-L}^{x_1-\varepsilon} \di x_2 \int_{x_2-L}^{x_2-\varepsilon} \di x_3  \int_{x_3-L}^{x_3-\varepsilon} \di x_4 \\ \Bigg( (a-1)^2 (x_1-x_2)^{a-2}(x_1-x_3)(x_1-x_4)^{a-1}(x_2-x_3)^{a-1}(x_2-x_4)(x_3-x_4)^{a-2} \\ - (x_1-x_2)^{a-1}(x_1-x_4)^{a-1}(x_2-x_3)^{a-1}(x_3-x_4)^{a-1} \\  + (a-1)^2 (x_1-x_2)^{a-1}(x_1-x_3)(x_1-x_4)^{a-2}(x_2-x_3)^{a-2}(x_2-x_4)(x_3-x_4)^{a-1} \Bigg)
\end{multline}

This integrand is too much coupled in its variables and not analytically computable in this form. But one can show using the identity
\begin{equation}
(x_1-x_2)(x_3-x_4)-(x_1-x_3)(x_2-x_4) + (x_1-x_4)(x_2-x_3)=0
\end{equation} 
that the integrand can be reexpressed as
\begin{multline}
\int \di x_1 \, e^{4\omega X^0} \int_{x_1-L}^{x_1-\varepsilon} \di x_2 \int_{x_2-L}^{x_2-\varepsilon} \di x_3  
\int_{x_3-L}^{x_3-\varepsilon} \di x_4 \\ \Bigg( (a-1)^2 (x_1-x_2)^{a-2}(x_2-x_3)^a (x_3-x_4)^{a-2}(x_1-x_4)^{a} \\ +\parent{2(a-1)^2 - 1} 
(x_1-x_2)^{a-1}(x_1-x_4)^{a-1}(x_2-x_3)^{a-1}(x_3-x_4)^{a-1} \\  + (a-1)^2 (x_1-x_2)^{a}(x_2-x_3)^{a-2}(x_3-x_4)^{a}(x_1-x_4)^{a-2} \Bigg)
\end{multline}

If we use the change of variable 
\begin{align}
x_2 = -L \delta_1 + x_1  \nonumber \\
x_3 = -L \delta_2 + x_2  \nonumber \\
x_4 = -L \delta_3 + x_3
\end{align}
the integral becomes: 
\begin{multline}
\int \di x_1 \, e^{4\omega X^0} \int_\eta^{1} \di \delta_1 \int_\eta^{1} \di \delta_2 \int_\eta^{1} \di \delta_3 \\ 
\Bigg( (a-1)^2 \delta_1^{a-2}\delta_2^a \delta_3^{a-2}(\delta_1+\delta_2+\delta_3)^{a} + (a-1)^2 \delta_1^{a}\delta_2^{a-2} 
\delta_3^{a}(\delta_1+\delta_2+\delta_3)^{a-2} \\ +\parent{2(a-1)^2 - 1} \delta_1^{a-1}\delta_2^{a-1} \delta_3^{a-1}(\delta_1+\delta_2+\delta_3)^{a-1} \Bigg)
\end{multline}

It is possible to extract the divergences by analytic integration but we need to be careful since we will need at some point to commute the integrals and 
sums. For this reason, the $z$-argument in the $\,_2 \text{F}_1(a,b,c,z)$ should satisfy $\module{z}<1$. 

We will not develop the whole computation, but give as  an example  one of the three integrals. Let us study the following one: 
\begin{equation} \label{eq:third_int}
\int_\eta^{1} \di \delta_1 \int_\eta^{1} \di \delta_2 \int_\eta^{1} \di \delta_3 ~  \delta_1^{a}\delta_2^{a-2} \delta_3^{a}(\delta_1+\delta_2+\delta_3)^{a-2}
\end{equation}

Integration of $\delta_3$ imposes to separate the domain of integration in three parts: 
\begin{align}
& \delta_1 + \delta_2 > 1  ~ \text{and} ~ \delta_3 \in [\eta ; 1] < \delta_1 + \delta_2 \nonumber \\
& \delta_1 + \delta_2 < 1  ~ \text{and} ~ \delta_3 \in [\eta ; \delta_1 + \delta_2] < \delta_1 + \delta_2 \nonumber \\
& \delta_1 + \delta_2 < 1  ~ \text{and} ~ \delta_3 \in [\delta_1 + \delta_2 ; 1] > \delta_1 + \delta_2
\end{align}
This makes three integrals: 
\begin{align}
& I_1 = \int_\eta^{1} \di \delta_1 \int_{1-\delta_1}^1 \di \delta_2 \int_\eta^{1} \di \delta_3 ~  \delta_1^{a}\delta_2^{a-2} \delta_3^{a}(\delta_1+\delta_2)^{a-2}(1+\frac{\delta_3}{\delta_1+\delta_2})^{a-2}  \nonumber \\
& I_2 = \int_\eta^{1} \di \delta_1 \int_\eta^{1-\delta_1} \di \delta_2 \int_\eta^{\delta_1 + \delta_2} \di \delta_3 ~  \delta_1^{a}\delta_2^{a-2} \delta_3^{a}(\delta_1+\delta_2)^{a-2}(1+\frac{\delta_3}{\delta_1+\delta_2})^{a-2}  \nonumber \\
& I_3 = \int_\eta^{1} \di \delta_1 \int_\eta^{1-\delta_1} \di \delta_2 \int_{\delta_1 + \delta_2}^1 \di \delta_3 ~  \delta_1^{a}\delta_2^{a-2} \delta_3^{2a-2}(1+\frac{\delta_1+\delta_2}{\delta_3})^{a-2}
\end{align} 
which integrate to:
\begin{align}
& I_1 = \int_\eta^{1} \di \delta_1 \int_{1-\delta_1}^1 \di \delta_2  ~  \delta_1^{a}\delta_2^{a-2} (\delta_1+\delta_2)^{a-2} \Bigg[ \frac{\delta_3^{a+1}}{a+1} \,_2\text{F}_1(2-a,a+1,a+2,-\frac{\delta_3}{\delta_1+\delta_2}) \Bigg]_\eta^1  \nonumber \\
& I_2 = \int_\eta^{1} \di \delta_1 \int_\eta^{1-\delta_1} \di \delta_2  ~  \delta_1^{a}\delta_2^{a-2} (\delta_1+\delta_2)^{a-2}\Bigg[ \frac{\delta_3^{a+1}}{a+1} \,_2\text{F}_1(2-a,a+1,a+2,-\frac{\delta_3}{\delta_1+\delta_2}) \Bigg]_\eta^{\delta_1+\delta_2}   \nonumber \\
& I_3 = \int_\eta^{1} \di \delta_1 \int_\eta^{1-\delta_1} \di \delta_2  ~  \delta_1^{a}\delta_2^{a-2} \Bigg[ \frac{\delta_3^{2a-1}}{2a-1} \,_2\text{F}_1(2-a,1-2a,2-2a,-\frac{\delta_1+\delta_2}{\delta_3}) \Bigg]_{\delta_1+\delta_2}^1
\end{align} 

We will not develop the computations for all the three integrals. Let us focus on the third, which is easier to present. The method is similar for the two other ones. 
\begin{multline}
I_3 = \int_\eta^{1} \di \delta_1 \int_\eta^{1-\delta_1} \di \delta_2  ~  \delta_1^{a}\delta_2^{a-2} \Bigg( \frac{1}{2a-1} \,_2\text{F}_1(2-a,1-2a,2-2a,-\delta_1-\delta_2) \\ 
- \frac{(\delta_1+\delta_2)^{2a-1}}{2a-1} \,_2\text{F}_1(2-a,1-2a,2-2a,-1) \Bigg)
\end{multline}
These are two different integrations to do.  We have: 
\begin{align}
& I_3^1 = \int_\eta^{1} \di \delta_1 \int_\eta^{1-\delta_1} \di \delta_2  ~  \delta_1^{a}\delta_2^{a-2} \frac{1}{2a-1} \,_2\text{F}_1(2-a,1-2a,2-2a,-\delta_1-\delta_2) \nonumber  \\
& I_3^2 = -\int_\eta^{1} \di \delta_1 \int_\eta^{1-\delta_1} \di \delta_2  ~ \frac{(\delta_1+\delta_2)^{2a-1}}{2a-1} \,_2\text{F}_1(2-a,1-2a,2-2a,-1)
\end{align}

Each of these separates again in three parts: 
\begin{align} \label{eq:dom_int_2}
&\delta_1 \in [\eta ; \frac{1}{2}] ~ \text{and} ~ \delta_2 \in [\eta ; \delta_1] \nonumber \\
&\delta_1 \in [\eta ; \frac{1}{2}] ~ \text{and} ~ \delta_2 \in [\delta_1 ; 1-\delta_1] \nonumber \\
&\delta_1 \in [\frac{1}{2};1] ~ \text{and} ~ \delta_2 \in [\eta ; 1-\delta_1] 
\end{align}

There is no known expression for the integration of $I_3^1$, but it is not much of a problem since we only want to extract divergences. 
Because $\module{\delta_1+\delta_2}<1$, one can express $\,_2 \text{F}_1$ as its series expansion given in~\refe{eq:series_F}. 
Since the series is convergent everywhere in the integration domain, we can commute the sum and the integral, such that
\begin{multline}
I_3^1 = - \sum_{n=0}^\infty \frac{\Gamma(a-1)}{\Gamma(a-1-n)\Gamma(1+n)(1-2a+n)} \\ 
\times \Bigg( \int_\eta^{1/2} \di \delta_1 \int_\eta^{\delta_1} \di \delta_2  ~  \delta_1^{a}\delta_2^{a-2} (\delta_1+\delta_2)^n 
+\int_\eta^{1/2} \di \delta_1 \int_{\delta_1}^{1-\delta_1} \di \delta_2  ~  \delta_1^{a}\delta_2^{a-2} (\delta_1+\delta_2)^n \\
+ \int_{1/2}^1 \di \delta_1 \int_{\eta}^{1-\delta_1} \di \delta_2  ~  \delta_1^{a}\delta_2^{a-2} (\delta_1+\delta_2)^n \Bigg)
\end{multline}
These integrals are very similar to the ones studied in appendix~\ref{annexe_CTT}. Following the method presented there, 
and with a careful power analysis in $\eta$, we can obtain: 
\begin{multline}
I_3^1 = -\sum_{n=0}^\infty \frac{\Gamma(a-1)}{\Gamma(a-1-n)\Gamma(1+n)(1-2a+n)} \\ 
\times \Bigg( -\frac{2^{-a-1-n} \eta^{a-1}}{(a+1+n)(a-1)} + o(1) + \frac{\parent{2^{-a-1-n} - 1} \eta^{a-1}}{(a+1+n)(a-1)} \Bigg)  \\
 = - \frac{\eta^{a-1}}{3a(a-1)}\parent{\frac{\,_2 \text{F}_1(2-a,1-2a,2-2a,-1)}{2a-1} + \frac{\,_2 \text{F}_1(2-a,a+1,a+2,-1)}{a+1}  }  + o(1)
\end{multline}  

The computation of $I_3^2$ is less difficult. With method of appendix~\ref{annexe_CTT} and~\refe{eq:dom_int_2}, it gives:
\begin{multline}
I_3^2 = \int_\eta^{1/2} \di \delta_1 ~ \delta_1^{4a-2}  \parent{\frac{\,_2 \text{F}_1(1-2a,a-1,a,-1)}{a-1} + \frac{\,_2 \text{F}_1(1-2a,2-3a,3-3a,-1)}{2-3a}} \\ \times \frac{\,_2\text{F}_1(2-a,1-2a,2-2a,-1)}{1-2a} \\ 
+\frac{\eta^{4a-1}}{3a(a-1)(4a-1)} \Bigg(3a \,_2 \text{F}_1(1-2a,a-1,a,-1)+(a-1) \,_2 \text{F}_1(1-2a,-3a,1-3a,-1)\Bigg) \\ \times \frac{\,_2\text{F}_1(2-a,1-2a,2-2a,-1)}{1-2a} \\ 
+ \frac{\eta^{a-1}}{3a} \frac{\,_2\text{F}_1(2-a,1-2a,2-2a,-1)}{a-1} + o(1)
\end{multline}

We do not integrate explicitely the first term so that the logarithm appears unambiguously at $a=1/4$. This has to be compared to the second term which does not become a 
logarithm, since it is finite at $a=1/4$. Indeed, for this precise value $a-1=-3a$ and one gets $\frac{\eta^{0}}{3a(a-1)}$. 

Finally, summing up $I_3^1$ with $I_3^2$, one obtains:
\begin{multline}
I_3 = \int_\eta^{1/2} \di \delta_1 ~ \delta_1^{4a-2}  \parent{\frac{\,_2 \text{F}_1(1-2a,a-1,a,-1)}{a-1} + \frac{\,_2 \text{F}_1(1-2a,2-3a,3-3a,-1)}{2-3a}} \\ \times \frac{\,_2\text{F}_1(2-a,1-2a,2-2a,-1)}{1-2a} \\ 
+\frac{\eta^{4a-1}}{3a(a-1)(4a-1)} \Bigg(3a \,_2 \text{F}_1(1-2a,a-1,a,-1)+(a-1) \,_2 \text{F}_1(1-2a,-3a,1-3a,-1)\Bigg) \\ \times \frac{\,_2\text{F}_1(2-a,1-2a,2-2a,-1)}{1-2a} \\ 
- \frac{\eta^{a-1}}{3a} \frac{\,_2\text{F}_1(2-a,a+1,a+2,-1)}{a+1} + o(1)
\end{multline}

Similarly one computes $I_1$ and $I_2$, for which we obtain: 
\begin{equation}
I_1 = o(1) 
\end{equation}
and 
\begin{multline}
I_2 = \int_\eta^{1/2} \di \delta_1 ~ \delta_1^{4a-2}  \parent{\frac{\,_2 \text{F}_1(1-2a,a-1,a,-1)}{a-1} + \frac{\,_2 \text{F}_1(1-2a,2-3a,3-3a,-1)}{2-3a}} \\ \times \frac{\,_2\text{F}_1(2-a,a+1,a+2,-1)}{a+1} \\ 
+ \frac{\eta^{4a-1}}{3a(a-1)(4a-1)} \Bigg(3a \,_2 \text{F}_1(1-2a,a-1,a,-1)+(a-1) \,_2 \text{F}_1(1-2a,-3a,1-3a,-1)\Bigg) \\ \times \frac{\,_2\text{F}_1(2-a,a+1,a+2,-1)}{a+1} \\
+ \eta^{4a-1} \sum _{n=0}^{\infty } \frac{\Gamma (a-1) }{\Gamma (n+1) \Gamma (a-n-1)(a+n+1)(3 a-n-2)  } \\ \left(\frac{\, _2 \text F_1(-2 a+n+1,-a+n+2;-2 a+n+2;-1)}{-2 a+n+1}+\frac{\, _2\text F_1(-2 a+n+3,-a+n+2;-2 a+n+4;-1)}{-2 a+n+3}\right) \\
- \frac{\eta^{a-1}}{3a} \frac{\,_2\text{F}_1(2-a,a+1,a+2,-1)}{a+1}  + o(1)
\end{multline}

One expresses then  the whole integral~\refe{eq:third_int} as
\begin{multline}
\int_\eta^{1} \di \delta_1 \int_\eta^{1} \di \delta_2 \int_\eta^{1} \di \delta_3 ~  \delta_1^{a}\delta_2^{a-2} \delta_3^{a}(\delta_1+\delta_2+\delta_3)^{a-2} \\ 
\sim \int_\eta^{1/2} \di \delta_1 ~ \delta_1^{4a-2}  \parent{\frac{\,_2 \text{F}_1(1-2a,a-1,a,-1)}{a-1} + \frac{\,_2 \text{F}_1(1-2a,2-3a,3-3a,-1)}{2-3a}} \\ \times \parent{ \frac{\,_2\text{F}_1(2-a,1-2a,2-2a,-1)}{1-2a} + \frac{\,_2\text{F}_1(2-a,a+1,a+2,-1)}{a+1}} \\
+ \eta^{4a-1} \Bigg[ \frac{1}{(4a-1)} \Bigg(\frac{\,_2 \text{F}_1(1-2a,a-1,a,-1)}{a-1} + \frac{\,_2 \text{F}_1(1-2a,-3a,1-3a,-1)}{3a} \Bigg) \\ \times \Bigg( \frac{\,_2\text{F}_1(2-a,1-2a,2-2a,-1)}{1-2a}  + \frac{\,_2\text{F}_1(2-a,a+1,a+2,-1)}{a+1} \Bigg) \\
+ \sum _{n=0}^{\infty } \frac{\Gamma (a-1) }{\Gamma (n+1) \Gamma (a-n-1)(a+n+1)(3 a-n-2)  } \Bigg(\frac{\, _2 \text F_1(-2 a+n+1,-a+n+2;-2 a+n+2;-1)}{-2 a+n+1}  \\ + \frac{\, _2 \text F_1(-2 a+n+3,-a+n+2;-2 a+n+4;-1)}{-2 a+n+3}\Bigg)\Bigg] \\
- 2\, \frac{\eta^{a-1}}{3a} \frac{\,_2\text{F}_1(2-a,a+1,a+2,-1)}{a+1}  + o(1)
\end{multline}
Similar techniques apply to the two other kinds of integrals.

\section{Partition function to eighth order}
\label{app:partition}

The disk partition function for the system,  unintegrated over the time-like zero modes, can be expressed as a series: 
\eqali{
Z(r|x_0) = \sum_{n=0}^\infty \parent{-\lambda^+\lambda^-e^{2\omega x^0}}^n I_{n}
}
with $I_{n}$ a coefficient that is equal to the sum time-ordered integrals that appear at order $n$ in the perturbative 
expansion. We can express it in a condensed form as : 
\eqali{
I_n &= \int [\underset{>}{dt}]_{2n} \left| \begin{array}{ccc} 1~3~5 &\ldots & 2n-1 \\ 2~4~6 &\ldots & 2n \end{array} \right|^{-4r^2}  
\sum_{perm~{\mathcal P}} (-1)^{P} \left|\begin{array}{c}  a_1~a_2 \\ a_3~a_4 \\ \ldots \\ a_{2n-1} ~ a_{2n}\end{array} \right| 
\parent{1-4r^2}^{\frac{n}{2}-\frac{1}{2} \sum_{i=1}^{n}(-1)^{a_{\scriptscriptstyle{2i-1}} - a_{\scriptscriptstyle{2i}} }} \nonumber \\ \label{eq:part_funct_no}
}
with the time-ordered measure 
\eqali{
[\underset{>}{dt}]_{2n} = \prod_{i=1}^{2n} \frac{dt_i}{2\pi} \prod_{i=1}^{2n-1} \Theta(t_i-t_{i+1}) 
}

We have also introduced convenient notations for the integrand, defined as:
\eqali{\label{eq:integnotations}
\left|\begin{array}{c}  a_1~a_2 \\ a_3~a_4 \\ \ldots \\ a_{2n-1} ~ a_{2n}\end{array} \right| = & \prod_{i=1}^{n}\prod_{j=2i+1}^{2n} S(a_{2i-1},a_j)S(a_{2i},a_j) 
\quad , \qquad \symform{i_1~i_2 \ldots i_p}{j_1 ~ j_2 \ldots j_n}=\prod_{\alpha=1}^p \prod_{a=1}^n S(i_\alpha,j_a) }
where $S(i,j)= \module{2\sin\frac{t_i-t_j}{2}}$. The sum in \refe{eq:part_funct_no} is done over all permutations within the set $\{1,2,3\ldots 2n\}$.\footnote{To be precise, we have 
$P(\{1,2,3\ldots 2n\})=\{a_1,a_2,a_3\ldots a_{2n}\}$.} Up to $n=2$, the partition function, for given $|r|<1/\sqrt 2$, reads: 
\eqali{
Z(r|x_0) & = 2 - 2\, \lambda^+\lambda^-e^{2\omega x^0} \int [\underset{>}{dt}]_{2} \symform{1}{2}^{-4r^2} \parent{1-4r^2} \nonumber \\  
             & + 2\parent{\lambda^+\lambda^-e^{2\omega x^0}}^2 \int [\underset{>}{dt}]_{4}  \symform{1~3}{2~4}^{-4r^2} 
             \parent{\parent{1-4r^2}^2 \symform{1~2}{3~4} - \symform{1~3}{2~4} +  \parent{1-4r^2}^2 \symform{1~4}{2~3}} \nonumber \\ 
             & + \ldots
\label{eq:4thordrer}}
The computation at second order in $T$, for $r \leq 1/2$, gives the result 
\eqali{
Z(r|x_0) = 2 - \frac{\Gamma(2-4r^2)}{\Gamma^2(1-2r^2)} \lambda^+\lambda^- e^{2\omega x_0} + 
\mathcal{O}\Big((\lambda^+ \lambda^-)^2\Big)   \label{eq:part_order_1}
}
where we used the Dyson integral~\cite{Forrester:2007}: 
\begin{equation}
\int_0^{2 \pi} \prod_{i=1}^n \frac{\di t_i}{2 \pi} \prod_{i<j}^n \module{e^{i \, t_i}-e^{i \, t_j}}^{2\alpha} = \frac{\Gamma(1 + n \alpha)}{\Gamma^n(1+\alpha)}
\end{equation}

We notice that the result~\refe{eq:part_order_1} is analytic in $r$ for all values below the critical distance $r_c = 1/\sqrt{2}$. The reason for this property 
should now be familiar to the reader. For $|r|<1/2$, the contact term vanishes, hence give no contribution to~(\ref{eq:part_order_1}). The value 
$r=1/2$ is particular. We see that the prefactor of the second order integral in~(\ref{eq:4thordrer}) vanishes; at the same time, the contact 
term gives a finite contribution, ensuring the continuity of the result in~(\ref{eq:part_order_1}). For any $1/2<|r|<r_c$, the second-order integral 
in~(\ref{eq:part_order_1}) is divergent. As we explained in subsec.~\ref{sec:contact_term}, the divergence is canceled by the contribution from the contact term, that 
appear in the worldsheet action~(\ref{eq:wscompaction}), where $\varepsilon$ is chosen to be the same as the short distance cutoff in~(\ref{eq:4thordrer}). The finite 
part that remains agrees precisely with~(\ref{eq:part_order_1}). Hence, the presence of the contact term gives (at least at this order) a continuous result all the 
way to the critical distance.

\subsection*{The r=1/2 case}
Finding the complete expression of the disk partition function at any $|r|<1/\sqrt 2$  seems to be out of reach, since integrals involve complicated highly coupled 
multidimensional integrals with path-ordering.   

For this reason we want to compute the partition function in the special case where $\omega=1/2$ which is tractable.  We recall that the 
perturbative expansion of the worldsheet action is given by~\refe{eq:action}: 
\eqna{
Z(r,\lambda^+,\lambda^-) &=& \corr{e^{-\delta S}} \nonumber \\
&=& \corr{e^{-\frac{ \lambda^+}{2\pi} \int 
d \hat t \, \Gam^+ \, T^+(\hat t) -\frac{ \lambda^-}{2\pi} \int d \hat t \, \Gam^- \, T^-(\hat t)}} \nonumber \\
&=& \sum_{n,p=0}^{\infty}(-1)^{n+p}\frac{(\lambda^+)^n}{n!}\frac{(\lambda^-)^p}{p!} \sum_{\text{perm} \pm}\int [d\hat t]_{n+p} \corr{\Gam^\pm (\hat t_1) 
\ldots \Gam^\pm(\hat t_{n+p})}\, \times\nonumber \\&&\	\phantom{aaaaaaaaaaaaaaaaaaaaaaaaaaaaaa} 
\times\,  \corr{T^\pm (\hat t_1) \ldots T^\pm (\hat t_{n+p})} 
}
with $n$ and $p$ of the same parity and $T^\pm=e^{\frac{\pm i \tilde {\mathbb X} + \omega {\mathbb X}^0}{2}}$.

Due to the Fermi multiplets correlators, the only non-vanishing terms are the ones which have as much $+$ as $-$. The correlator of 
the Fermi multiplets are easy to compute using Wick theorem and the Green function~\refe{eq:gam_gam}. 
It leads to one product of supersymmetric sign functions $\prod \hat \epsilon(2i,2i+1)$, which decomposes into a sum of $2 (n!)^2$ supersymmetric path orderings. 
One finds that these path orderings are all equivalent under permutations of $T^+$'s ($T^-$'s) with $T^+$'s ($T^-$'s) 
and permutations of integration variable. So one choose one path ordering, symbolically $\hat t_1>\hat t_2>\ldots>\hat t_{2n}$ 
multiplied by a factor $2 (n!)^2$. We should then compute: 
\eqna{
Z(r,\lambda^+,\lambda^-)&=& 2 \sum_{n=0}^{\infty} (\lambda^+\lambda^-)^n \int [\underset{>}{d\hat t}]_{2n} 
\corr{T^+ (\hat t_1)T^- (\hat t_2) \ldots T^-(\hat t_{2n})} \nonumber  \\ 
&=& 2 \sum_{n=0}^{\infty}(\lambda^+\lambda^-)^n e^{in x} \int [\underset{>}{d\hat t}]_{2n} \prod_{i<j}^n \hat S(2i,2j) \hat S(2i-1,2j-1) \nonumber \\
&=& 2 \sum_{n=0}^{\infty} (\lambda^+\lambda^-)^n e^{i n x} \; I_n 
} 
with $\hat S(i,j)=|2 \sin \frac{t_i-t_j}{2}|-\epsilon(i,j)\theta_i \theta_j$.

The computation of the integrals $I_n$ is as follows, using the notation of eq.~(\ref{eq:integnotations}). We have first
\eqna{
I_1 = -  \frac{1}{2\pi} \int [dt]_1 =
	 -\frac{1}{2\pi}
}
Then $I_2$ which is still easy
\eqna{
I_2  =\frac{1}{(2\pi)^2}\int \frac{[dt]_2}{2!} 
\symform{1}{2}  \symform{2}{1} - \int \frac{[dt]_{4}}{4!} 
	= \frac{1}{(2\pi)^2} - \frac{1}{4!} 
}
and
\begin{equation}
I_3 = \frac{1}{2\pi}\int \frac{[dt]_{5}}{5!} C_1^5 \symform{1}{2345}-\frac{1}{(2\pi)^3}\int \frac{[dt]_3}{3!} 
\symform{1}{23}  \symform{2}{13}\symform{3}{12} 
= \frac{2^{12}}{4!(2\pi)^5} -\frac{1}{(2\pi)^3} 
= \frac{16}{3\pi^5} - \frac{1}{8\pi^3}
\end{equation}

$I_4$ is a bit more complicated to compute, but in terms of integrals, we find: 
\eqna{
I_4 &=&  \int  [\underset{>}{d t}]_8  \; \Asymform{13}{57}\Asymform{24}{68}   - \frac{1}{(2\pi)^2} 
\int \frac{[dt]_6}{6!} \; C_2^6 \; \symform{1}{2}\symform{2}{1} \symform{12}{3456} \nonumber \\ && \quad + \frac{1}{(2\pi)^4} 
\int \frac{[dt]_4}{4!} \; \symform{1}{234}\symform{2}{134}\symform{3}{124}\symform{4}{123}  \nonumber \\
    &=& \frac{1}{1120}+\frac{143}{144 \pi ^6}-\frac{55}{192 \pi ^4}+\frac{13}{480 \pi ^2} -
    \parent{-\frac{1001}{2592 \pi ^6}-\frac{175}{432 \pi ^4}-\frac{1}{240 \pi ^2}} + \frac{1}{16 \pi^4} \nonumber \\
    &=& \frac{1}{1120}+\frac{3575}{2592 \pi ^6}+\frac{205}{1728 \pi ^4}+\frac{1}{32 \pi ^2} + \frac{1}{16 \pi^4}
}
where we introduced the totally antisymmetric form : 
\eqna{
\Asymform{ab\ldots}{cd\ldots}&=& \sum_P (-1)^P \asymform{p(a)p(b)\ldots}{p(c)p(d)\ldots} \nonumber \\
							&=& \asymform{ab\ldots}{cd\ldots} - 
							\asymform{ac\ldots}{bd\ldots} + \asymform{ad\ldots}{bc\ldots} + \ldots \nonumber \\
}
with the partially anti-symmetric form :
\eqna{
\asymform{abc\ldots}{def\ldots} = \epsilon(a,b)\epsilon(a,c)\epsilon(b,c)\times 
\ldots \times \epsilon(d,e)\epsilon(d,f)\epsilon(e,f)\times \ldots \times \symform{abc \ldots}{def \ldots}
}
The bigger $n$ is, the more complicated is the corresponding term in the partition function. This is because more and more 
contribution of the contact term appear and that the path ordering can't be always removed. For the special value $r=1/2$ the contact term has 
indeed a non-zero, but finite contribution to  the final result.

We end up with the following expansion. The terms coming from pure 'non-contact' contributions are underlined: 
\eqna{
\frac{Z(x)}{2} &=& 1 - \lambda^+\lambda^- \frac{e^{i x}}{2\pi} + \frac{\parent{\lambda^+\lambda^-}^2}{4\pi^2} 
e^{2ix}\parent{1 - \underline{\frac{\pi^2}{6}}} -  \frac{\parent{\lambda^+\lambda^-}^3}{8\pi^3} e^{3ix} \parent{1 - \frac{128}{3 \pi^2}} 
\nonumber \\ && + \frac{\parent{\lambda^+\lambda^-}^4}{16\pi^4} e^{4ix} \parent{1+\frac{175}{27}+\frac{1001}{162 \pi ^2}+\frac{\pi ^2}{15}
\underline{-\frac{55}{12}+\frac{143}{9 \pi ^2}+\frac{13 \pi ^2}{30}+\frac{\pi ^4}{70}}} \ldots \nonumber \\
}
where we recognize the trivial expansion : 
\eqna{
1 - \lambda^+\lambda^- \frac{e^{i x}}{2\pi} + \parent{\lambda^+\lambda^-}^2 \frac{ e^{2ix}}{4\pi^2} 
- \parent{\lambda^+\lambda^-}^3 \frac{ e^{3ix}}{8\pi^3} + \ldots = \frac{1}{1+ \frac{\lambda^+\lambda^-}{2\pi}e^{i x}} \label{eq:triv_exp}
}

In fact, this factorization is exact to all orders; by looking at the integrals $I_n$, one can see that the maximal contact term is always present and has a  
standard form, which we recognize as a Vandermonde determinant. 

The remaining terms should come from a non-trivial function that multiplies~\refe{eq:triv_exp}: 
\eqna{
Z(x) &=& \frac{2}{1+ \frac{\lambda^+\lambda^-}{2\pi}e^{i x}} \left(1- \parent{\frac{\lambda^+\lambda^-}{2\pi}}^2 \frac{\pi^2}{6}e^{2ix} + 
\parent{\frac{128}{3\pi^2}-\frac{\pi^2}{6}}\parent{\frac{\lambda^+\lambda^-}{2\pi}}^3 e^{3ix} \right. \nonumber \\ && \left. + 
\parent{\frac{205}{108}+\frac{10487}{162 \pi ^2}+\frac{\pi ^2}{2}+\frac{\pi ^4}{70}}\parent{\frac{\lambda^+\lambda^-}{2\pi}}^4 e^{4ix} +\ldots \right)
}
This doesn't seem to come from the Taylor expansion of a simple expression.

\end{document}